\documentclass[fleqn,usenatbib]{mnras}

\usepackage{newtxtext,newtxmath}

\usepackage[T1]{fontenc}
\usepackage{scalerel}
\DeclareRobustCommand{\VAN}[3]{#2}
\let\VANthebibliography\thebibliography
\def\thebibliography{\DeclareRobustCommand{\VAN}[3]{##3}\VANthebibliography}

\usepackage{graphicx}	
\usepackage{amsmath}	
\usepackage{algorithm}
\usepackage{algorithmic}

\title[Linearized radial velocity extraction]{A linearized approach to radial velocity extraction}

\author[S. Shahaf \& B. Zackay]{
S. Shahaf$^{1}$\thanks{E-mail: sahahr.shahaf@weizmann.ac.il}
and B. Zackay$^{1}$
\\
$^{1}$Department of Particle Physics and Astrophysics, Weizmann Institute of Science, Rehovot 7610001, Israel\\
}

\date{Accepted XXX. Received YYY; in original form ZZZ}

\pubyear{2022}

\begin{document}
\label{firstpage}
\pagerange{\pageref{firstpage}--\pageref{lastpage}}
\maketitle

\begin{abstract}
High-precision radial velocity (RV) measurements are crucial for exoplanet detection and characterisation. Efforts to achieve ${\sim}10$ cm\,s\textsuperscript{-1} precision have been made over the recent decades, with significant advancements in instrumentation, data reduction techniques, and statistical inference methods. However, despite these efforts, RV precision is currently limited to ${\sim}50$ cm\,s\textsuperscript{-1}. This value exceeds state-of-the-art spectrographs' expected instrumental noise floor and is mainly attributed to RV signals induced by stellar variability. In this work, we propose a factorisation method to overcome this limitation. The factorisation is particularly suitable for controlling the effect of localised changes in the stellar emission profile, assuming some smooth function of a few astrophysical parameters governs them. We use short-time Fourier transforms (STFT) to infer the RV in a procedure equivalent to least-squares minimisation in the wavelength domain and demonstrate the effectiveness of our method in treating arbitrary temperature fluctuations on the star's surface. The proposed prescription can be naturally generalised to account for other effects, either intrinsic to the star, such as magnetic fields, or extrinsic to it, such as telluric contamination. As a proof-of-concept, we empirically derive a set of factorisation terms describing the Solar centre-to-limb variation and apply them to a set of realistic SOAP-GPU spectral simulations. We discuss the method's capability to mitigate variability-induced RV signals and its potential extensions to serve as a tomographic tool. 
\end{abstract}




\section{Introduction}
\label{sec: introduction}
%

Extracting radial velocities (RV) from a time series of spectroscopic measurements is a long-withstanding challenge of observational astronomy. A few decades ago, when studying a large sample of galaxies, \citet{tonry79} used the cross-correlation function (CCF) between the observed spectra and a set of theoretical templates to estimate the RV of each object. The statistical equivalence between CCF maximisation and least-squares minimisation provided the formal justification for their approach, while its computational efficiency made it technically appealing (see \citealt{brault71}, who advocated for using fast Fourier transform in astronomy). 

Indeed, CCF-based techniques have been a standard practice of RV extraction for many years since \citep[e.g.,][]{ zucker94, pepe02, lovis10}. An important assumption underlying this method is that the chosen theoretical template accurately describes the observed data. Deviations from the expected spectral shape, in principle, violate this assumption. Some techniques were consequentially developed to estimate RV based on empirical models derived from the data itself \citep[e.g.,][]{zucker06, zechmeister18, rajpaul20}. In practice, whether derived from theoretical or empirical models, the RV inaccuracy induced by modelling errors was often small compared to the overall noise budget.

The search for planets beyond our solar system has shifted these tides. The required precision to detect an Earth-like planet orbiting a Sun-like star is ${\sim} 10\,\,{\rm cm}\,{\rm s}^{-1}$. Although state-of-the-art spectrographs are expected to achieve this precision in the foreseeable future, it has become increasingly clear that stellar variability hinders the efforts to detect and characterise such small signals (e.g., \citealt{fischer16, crass21,newman23, hara23}). Magnetically active regions on the stellar surface, for example, deform the shapes of the spectral lines and obscure the Doppler shift induced by the star's reflex motion \citep[e.g.,][]{figueira13,haywood15, plavchan15, luhn22}.  
Stellar variability can induce a spectroscopic signal, which modulates with stellar rotation or activity cycles. As a result, if some periodic RV modulation is detected, one faces the challenge of ascribing it to the star's reflex motion rather than spectral changes due to the presence of active regions on its surface (e.g., \citealt{lubin21, john22, john23}). Furthermore, even if an actual planet exists in the system, stellar variability might induce correlated noise on the resulting RV time series, impacting the ability to accurately measure its mass \citep{luhn23}.

Several ways to mitigate the influence of stellar activity have been developed in recent years (see the methodological overview by \citealt{zhao22} and \citealt{hara23} for a comprehensive statistical review). These techniques share a common objective and are often not mutually exclusive; however, they differ in their approach to the problem. For example, some aim to mitigate the effect of line-profile deformation by analysing the spectrum or its CCF as a whole \citep[e.g.,][]{davis17,cameron21, cretignier22}, while others focus on individual spectral lines \citep{cretignier20, artigau22, al-moulla22b, al-moulla22}. Some use activity indicators to quantify the contribution of stellar activity to each measurement \citep[e.g.,][]{milbourne19, haywood22}. In contrast, others use generalised noise models to post-priory estimate the activity-related contribution to the RV \citep[e.g.,][]{rajpaul15, barragan19, zicher22}. Alternative analysis frameworks, such as nonparametric statistical tests for periodicity detection \citep[e.g.,][]{hara17, binnenfeld20, binnenfeld22, hara22} and advanced machine learning algorithms \citep{beurs22, gully22}, are also being developed.

This work describes a new method to derive the RV in cases where \textit{the local emission pattern on the star's surface can be represented as a linear combination of known functions}. Our method relies on short-time Fourier transform (e.g., \citealt{grochenig01}), often used in audio signal processing. This work aims to present the mathematical and statistical framework underlying our method and to provide several examples to demonstrate its capabilities and design. Section~\ref{sec: math} presents the mathematical formulation; Section~{\ref{sec: RV ext}} describes the statistical framework; Section~\ref{sec: numerical experiments} shows the results from our numerical experiments; and Section~\ref{sec: discussion} summarises the work and presents prospects for future study.

\section{Mathematical formulation} 
\label{sec: math}

The measured spectrum of a star is an integrated emission profile summed over all visible points on its luminous disk. Therefore, it does not simply represent the star's local photospheric emission pattern but rather some averaged profile that reflects, for example, the stellar rotation rate and the finite resolution of the instrument. 

Some properties, such as stellar rotation and instrumental resolution, can be modelled as a convolution of the local stellar spectrum with some broadening kernel \citep[e.g.,][]{gray92}. Other phenomena, like star spots, locally change the emitted spectrum and consequently alter the shape of the disk-integrated profile \citep[e.g.,][]{dumusque14, zhao23}. Changes to the observed spectrum are not necessarily due to processes intrinsic to the astrophysical source, as telluric contamination or calibration errors can also manifest as deviations from the expected spectral shape \citep[e.g.][]{cretignier21, wang22, latouf22}.

This section provides the mathematical formulation we use to describe the spectrum. Its main goal is to describe how local changes to the emission pattern propagate to the final disk-integrated measurement. Active regions on the surface of the star can produce, in some cases, patches that are colder than their surrounding regions and manifest as star spots. Therefore, as a demonstration, we apply this formalism to describe a situation in which the stellar effective temperature can vary slightly from one point on the star to the other.

%
\subsection{Factoring stellar spectra}
\label{sec: expansion}
Consider a spectrum, $S(\lambda)$, sampled over a logarithmically uniform wavelength grid, such that the difference between two consecutive points is equivalent to some constant shift in velocity \citep[e.g.,][]{tonry79}.\footnote{We denote our independent variable as $\lambda$. This choice of notation aims to avoid confusion with other velocity-related terms. We emphasise that the $\lambda$-axis is a linear scale with units of velocity and resolution $\delta \lambda$. A typical size for the sampling resolution is $100-1000$ ${\rm m}\,{\rm s}^{-1}$, significantly larger than the amplitude of the RV modulations we seek.}  

We start by describing the emission per unit area from some particular position on the star's luminous disk. The position of the emitting area element is denoted by a two-dimensional position vector, $\mathbfit{r}$. As discussed earlier, the properties of the emission pattern may have several contributing factors. We assume to have some functional model describing each factor and that a linear combination of these functions can reproduce, or at least approximate, the emission pattern at each location on the stellar surface (for a practical example, see Section~\ref{sec: Teff expansion} below). More explicitly, we assume that the local emission profile, $f$, can be described as
\begin{equation}
    \label{eq: f rest frame}
    f(\lambda_{\rm e}; \, \mathbfit{r}) = 
    u_{_0}({\mathbfit{r}}) \, f^{{\scaleto{(0)}{5pt}}}(\lambda_{\rm e}) + 
    u_{_1}({\mathbfit{r}}) \, f^{{\scaleto{(1)}{5pt}}}(\lambda_{\rm e}) +  
    u_{_2}({\mathbfit{r}}) \, f^{{\scaleto{(2)}{5pt}}}(\lambda_{\rm e}) + ... \, ,
\end{equation}
where $\lambda_{\rm e}$ is the emitted wavelength at the rest frame of the area element.

A practical example of the decomposition in equation~(\ref{eq: f rest frame}) is a Taylor expansion of the photospheric model, which accounts for small local changes in the stellar parameters. We demonstrate such an approximation in Section~\ref{sec: Teff expansion}, where we take $u_{_k}(\mathbfit{r})$ to be a local temperature deviation,  $\Delta T (\mathbfit{r})$. However, to emphasise the generality of the decomposition, we keep the principal spectra, $f^{{\scaleto{(k)}{5pt}}}$, loosely defined at this point. We only assume that a linear combination of a few or a few dozen such spectra can describe the emission pattern of the star to sufficiently high precision. 

Different parts on the star's surface move with different velocities with respect to the observer due to the star's rotation. As a result, the observed emission from each area element is shifted by some $\Delta \lambda(\mathbfit{r})$. Using the sifting property of Dirac's delta function, the observed emission pattern can be written as
\begin{equation}
    \label{eq: f observer}
     f_{\rm o}(\lambda; \, \mathbfit{r}) \equiv \int \delta\big(\lambda'-\lambda + \Delta\lambda({\mathbfit{r}})\big) f(\lambda'; \, \mathbfit{r}) \, d\lambda' ,
\end{equation}
where $\lambda$ is the wavelength in the observer's frame, and $\Delta \lambda(\mathbfit{r})=\lambda-\lambda_{\rm e}$, is a function of the position on the star's disk.

So far, we have described the emission per unit area from some particular position on the face of the star. Practically, we observe the integrated contribution from the entire disk. Each area element, $\delta A$, contributes  $f_{\rm o}(\lambda;\, \mathbfit{r}) \delta A$ to the overall flux. To obtain the observed spectrum, we integrate all area elements,
\begin{equation}
    \label{eq: S int 1}
     S(\lambda)  = \iint_{_A} f_{\rm o}(\lambda; \,\mathbfit{r}) \,  dA \, .
\end{equation}
This double integral is taken over the projection of the star's luminous disk onto the plane of the sky, summing all its visible parts. 

The general integral expression in equation~(\ref{eq: S int 1}) can be broken down to a sum that represents the contributions of the functional terms we used to describe the local emission pattern in equation~(\ref{eq: f rest frame}). To do so, we plug equations~(\ref{eq: f rest frame}-\ref{eq: f observer}) into equation~({\ref{eq: S int 1}), and obtain
\begin{equation}
\label{eq: flux int}
     S(\lambda) 
    =\, \sum_{k} \, \bigg[ \iiint \delta\big(\lambda'-\lambda+\Delta\lambda({\mathbfit{r}})\big) u_{_k}(\mathbfit{r}) f^{{\scaleto{(k)}{5pt}}}(\lambda') dAd\lambda' \bigg].
\end{equation}
Each term in this series is now a triple integral. Two coordinates are positional, as the integral sums over the different area elements on the plane of the sky. The third integral is inherited from equation~(\ref{eq: f observer}) and represents the Doppler shift of each area element.

We now wish to simplify the expression in equation~(\ref{eq: flux int}). To do so, we define the $k$\textsuperscript{th} principal kernel, $U^{{\scaleto{(k)}{5pt}}}$, as the result of the integration over the plane of the sky,
\begin{equation}
    \label{eq: U integral def}
    U^{{\scaleto{(k)}{5pt}}}(\lambda)\equiv\iint_{_A}u_k(\mathbfit{r})\delta\big(\lambda + \Delta \lambda(\mathbfit{r})\big) dA \, .
\end{equation}
Using the definition above, the $k$\textsuperscript{th} term in the series shown in equation~(\ref{eq: flux int}) can be written as
 \begin{equation}
 \label{eq: S^k}
\begin{aligned}
    S^{{\scaleto{(k)}{5pt}}}(\lambda) = \int  U^{{\scaleto{(k)}{5pt}}}(\lambda'-\lambda) f^{{\scaleto{(k)}{5pt}}}(\lambda') d\lambda' \equiv   \left(U^{{\scaleto{(k)}{5pt}}}  \circledast f^{{\scaleto{(k)}{5pt}}}\right)(\lambda) \, .
\end{aligned}
\end{equation}
The integral equation above is a simple convolution of the $k$\textsuperscript{th} principal kernel with its corresponding principal spectra, taken in the $\lambda$-axis. To represent this convolution, we use the $\circledast$ symbol. 
 
Using equation~(\ref{eq: S^k}), we rewrite the expression for the disk-integrated spectrum as a sum of convolutions,
\begin{equation}
    \label{eq: S taylor conv}
    S(\lambda) = 
    U^{{\scaleto{(0)}{5pt}}}\circledast f^{{\scaleto{(0)}{5pt}}}(\lambda) + 
    U^{{\scaleto{(1)}{5pt}}}\circledast f^{{\scaleto{(1)}{5pt}}}(\lambda) + ... \,.
\end{equation}
The above-mentioned principal spectra, $f^{{\scaleto{(k)}{5pt}}}$, represent the expansion terms of the local emission pattern. This pattern is weighted, arranged and shifted by the principal kernels,  $U^{{\scaleto{(k)}{5pt}}}$. 
Equation~(\ref{eq: S taylor conv}) attains a simple form in the Fourier domain. This is because, by the convolution theorem, it is represented as a sum of products,
\begin{equation}
    \label{eq: S taylor fourier}
    \tilde{S}(\zeta) = 
    \tilde{U}^{{\scaleto{(0)}{5pt}}}\, \tilde{f}^{{\scaleto{(0)}{5pt}}}(\zeta) + 
    \tilde{U}^{{\scaleto{(1)}{5pt}}}\, \tilde{f}^{{\scaleto{(1)}{5pt}}}(\zeta) + ... \,  , 
\end{equation}
where the tilde sign indicates the Fourier transform (FT).

\begin{figure}
        \centering
        \includegraphics[width=0.975\columnwidth]{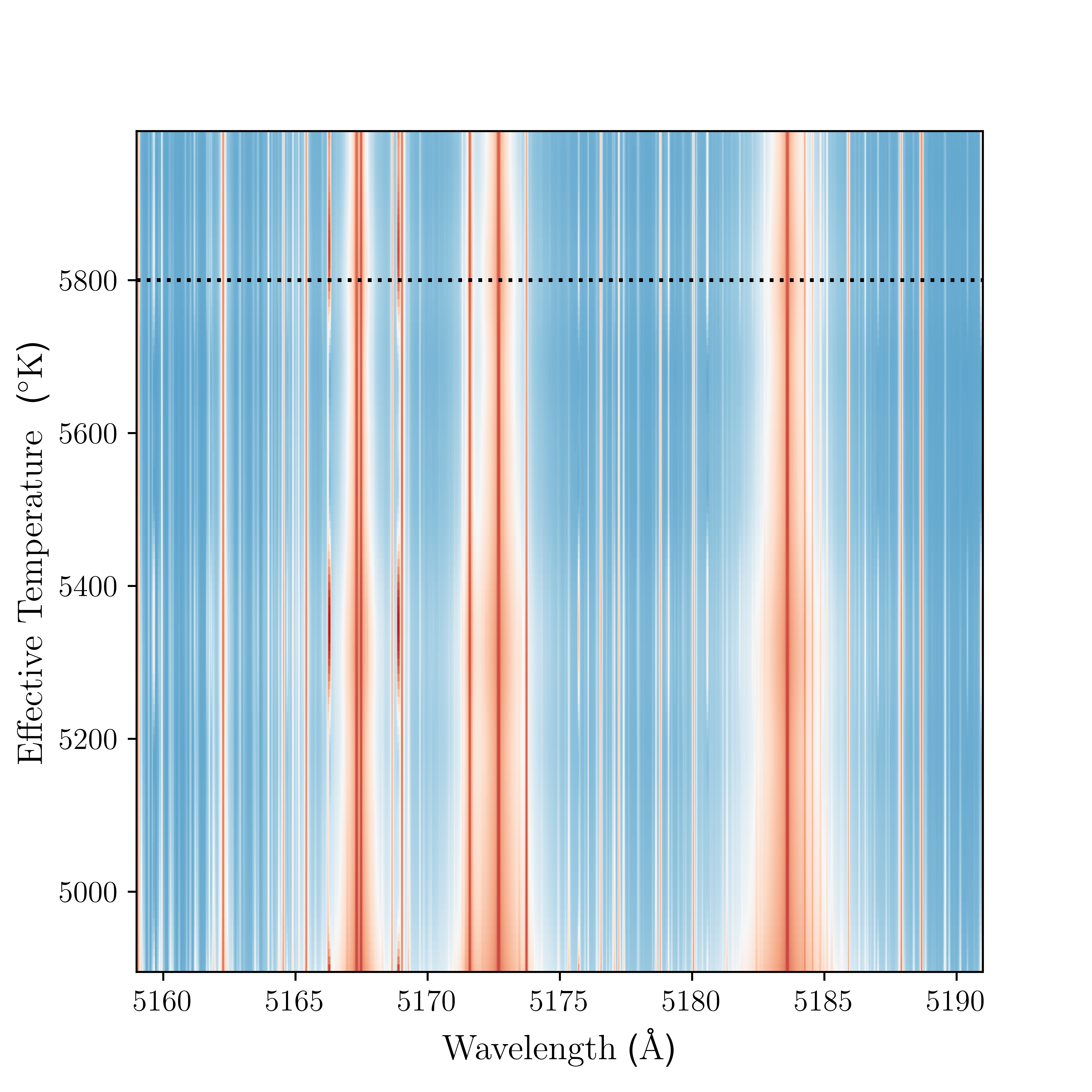} %
        \caption{Interpolated PHOENIX models, sampled around the MgB lines. The colours illustrate the continuum-normalized emission, where blue and red represent the high and low flux levels, respectively. The horizontal axis shows the wavelength grid, and the vertical one shows the effective temperature of the model. The dotted black line references the selected temperature at which we estimated the flux and its temperature derivatives (see Section~\ref{sec: Teff expansion} and Figure~\ref{fig: teff derivatives}).}
    \label{fig: teff map}
\end{figure}
%
\begin{figure}
        \centering 
        \includegraphics[width=0.975\columnwidth]{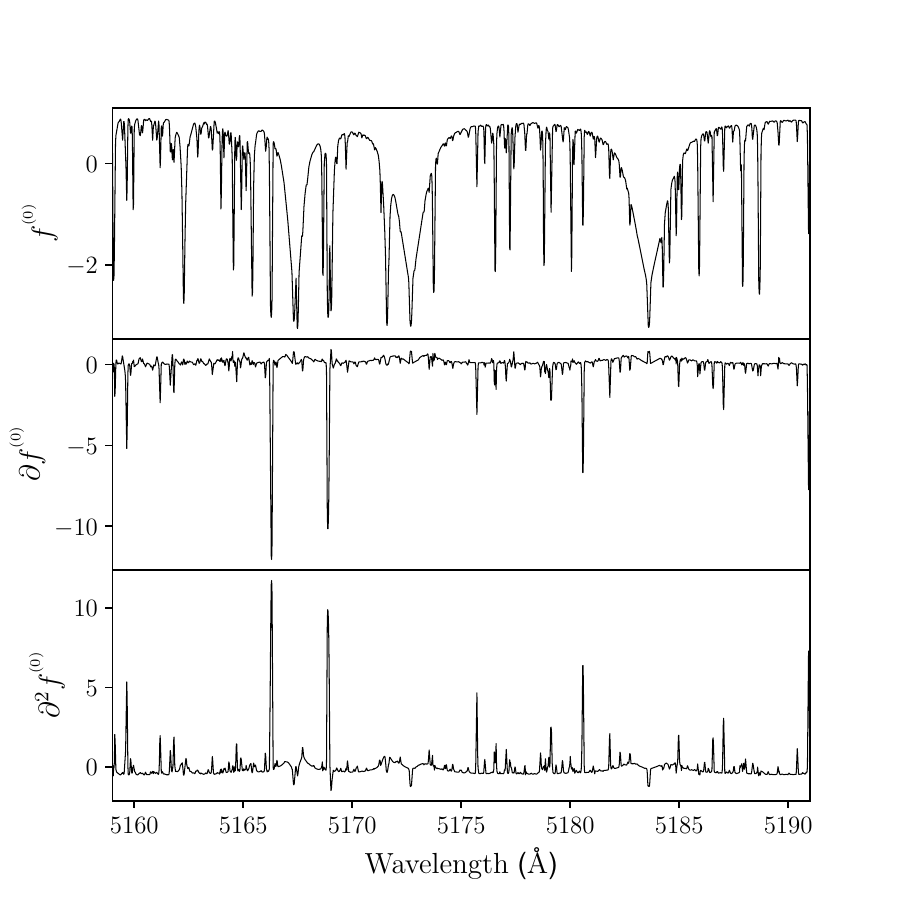} %
        \caption{Mean-subtracted and normalized sections of principal spectra used for the linearized photospheric models in Section~\ref{sec: Teff expansion}. Top panel ---  the zeroth-order principal spectra, $f^{{\scaleto{(0)}{5pt}}}$, a Sun-like model taken along the horizontal dotted lines in Figure 1. Middle panel --- the first-order principal spectra, $f^{{\scaleto{(1)}{5pt}}}$, taken as the derivative of the model with respect to the effective temperature. Bottom panel --- the second-order principal spectra, $f^{{\scaleto{(2)}{5pt}}}$, taken as the second derivative of the model with respect to the effective temperature.}
    \label{fig: teff derivatives}
\end{figure}

\subsection{Linearized photospheric models}
\label{sec: Teff expansion}

In the previous section, we assumed that the rest frame emission from any single point of the star is a linear combination of principal spectra. We now show that this is a natural consequence of the continuous manner in which the spectral properties change with respect to the star's physical properties. 

Although the stellar spectrum contains many different features, only a few properties are required to control its shape. These include, for example, stellar convective flows, effective temperature, pressure and metallicity. We construct a linear model using a simple Taylor expansion per wavelength, taken around the estimated stellar parameters. While changes to some parameters affect the spectrum very linearly, inducing only minor changes to it, others could cause nonlinear effects and require several terms for the expansion to be efficient (e.g., spectral line appearance or disappearance due to local temperature changes). %

The effectiveness of the linear decomposition is demonstrated by modeling local effective temperature ($T_{\rm eff}$) variations as a Taylor series. 
We define the first term in the expansion,  $f^{{\scaleto{(0)}{5pt}}}_{_T}$, to be the photospheric model at some reference temperature $T_0$. The high-order terms in our expansion are, therefore, given by 
\begin{equation}
    f^{{\scaleto{(k)}{5pt}}}_{_T} = \frac{\partial^{k} f^{{\scaleto{(0)}{5pt}}}_{_T}}{\partial T_{\rm eff}^{k}}\bigg|_{\scaleto{T_0}{5pt}}\!.
\end{equation}
Figure~\ref{fig: teff map} illustrates how the local emission pattern of a star changes as a function of $T_{\rm eff}$. The continuum-normalized photospheric models underlying this figure were generated by PHOENIX \citep{husser13}, and obtained from POLLUX\footnote{Publically available via this \hyperlink{http://npollux.lupm.univ-montp2.fr/}{link}.} database \citep{palacios10}. The figure shows how some lines become deeper or shallower, as the red and light-blue colours indicate. 
Figure~\ref{fig: teff derivatives} shows the first three terms we use to describe a Sun-like photospheric model.

Using the temperature expansion terms, equation~(\ref{eq: f rest frame}) for the local emission pattern can be written, up to second order, as 
\begin{equation}
    \label{eq: f rest frame Teff}
    f(\lambda_{\rm e}; \, \mathbfit{r}) \simeq
    f^{{\scaleto{(0)}{5pt}}}_{_T}(\lambda_{\rm e}) + 
    \Delta T({\mathbfit{r}}) \, f^{{\scaleto{(1)}{5pt}}}_{_T}(\lambda_{\rm e}) + 
    \Delta T({\mathbfit{r}})^2 \, f^{{\scaleto{(2)}{5pt}}}_{_T}(\lambda_{\rm e})\, ,
\end{equation}
where $\Delta T({\mathbfit{r}})$ is the local temperature deviation. By plugging equation~(\ref{eq: f rest frame Teff}) into equation~(\ref{eq: S taylor fourier}), we get an approximated prescription for the spectrum,
\begin{equation}
    \label{eq: S expansion Teff}
    \tilde{S}(\zeta)\simeq 
    \tilde{U}^{{\scaleto{(0)}{5pt}}} \tilde{f}^{{\scaleto{(0)}{5pt}}}_{_T}(\zeta) + \tilde{U}^{{\scaleto{(1)}{5pt}}} \tilde{f}^{{\scaleto{(1)}{5pt}}}_{_T}(\zeta) + \tilde{U}^{{\scaleto{(2)}{5pt}}} \tilde{f}^{{\scaleto{(2)}{5pt}}}_{_T}(\zeta) \, . 
\end{equation}

Using the equation above, we can draw some qualitative conclusions about the expected shape of the kernels. Since the first term in the expansion represents the fiducial model of the star, its expected shape will contain the instrumental line spread function, $\tilde{K}_{\rm inst}$, and the rotational broadening kernel, $\tilde{K}_{\rm rot}$. Therefore, 
\begin{equation}
    \label{eq: U0}
    \tilde{U}^{{\scaleto{(0)}{5pt}}}(\zeta) = \exp{\big(-i2\pi \Delta v \zeta\big)} \, \,\tilde{K}_{\rm inst}\big(\zeta ; \,\mathcal{R}\big)  \,\, \tilde{K}_{\rm rot}\big(\zeta ; \, v_{\rm rot}, \epsilon\big) \, , 
\end{equation}
where $\Delta v$, $v_{\rm rot}$, $\epsilon$ and $\mathcal{R}$ represent the star's RV, rotational velocity projection on our line of sight, linear limb-darkening coefficient and the instrumental resolution, respectively. In Appendix~\ref{app: broadening kernels} we provide the explicit expressions for $\tilde{K}_{\rm inst}$ and $\tilde{K}_{\rm rot}$, for completeness. 

To clarify the role of the kernels, consider the simplest case, in which the temperature deviation is localized to one particular location on the surface of the star, which moves with some velocity $v_{\rm spot}$ with respect to its centre-of-mass velocity. In this case, the high-order kernels will be limited by the instrumental resolution, hence
\begin{equation} 
\label{eq: Uk}
\tilde{U}^{{\scaleto{(k)}{5pt}}}(\zeta)\propto \exp{\big(-i2\pi(\Delta v + v_{\rm spot}) \zeta\big)} \,\, \tilde{K}_{\rm inst}\big(\zeta; \,\mathcal{R}\big)\, ,
\end{equation}
with some scaling factor that depends on $\Delta T$ at the position of the spot (see the simulations in Section~\ref{sec: moving spot simulation} for a demonstration). We emphasize that the localization of the temperature fluctuation is neither an assumption nor a requirement of our method, as shown in Section~\ref{sec: several spots}. For simplicity, we do not discuss the velocity induced by convective blueshift inhibition at the moment (but see Section~\ref{sec: soap simulation}).

\section{RV extraction}
\label{sec: RV ext}
Each additive term in equation~(\ref{eq: S taylor conv}) is a convolution of two functions: the principal spectra, $f^{{\scaleto{(k)}{5pt}}}$, which are assumed to be known, and their corresponding kernels, $U^{{\scaleto{(k)}{5pt}}}$, which we wish to infer from the spectrum. These two types of functions are fundamentally different:

First, the principal spectra can have any general form in the $\lambda$-axis; their corresponding convolution kernels, however, span a limited range and coincide with zero away from the origin (i.e., have compact support). For instance, the rotational broadening kernel cannot blend spectral regions separated by more than twice the star's rotational velocity. Similarly, the line spread function effectively drops to zero for wavelengths separated by more than a few times $c/\mathcal{R}$. In Section~\ref{sec: compact support}, we use this property to show that, provided with a set of ${\sim}10$ principal spectra, one could infer the shapes of the kernels based on a single spectrum.

Second, the principal spectra in our formulation describe all the chromatic changes in the spectrum. Therefore, if we apply equation~(\ref{eq: S taylor fourier}) to a small spectrum segment, the principal spectra will change, while their corresponding kernels will not. Qualitatively, one could think about this step as analysing data obtained by an echelle spectrograph order-by-order: the rotational broadening profile, for example, is assumed to be the same for all orders even though different orders contain a different set of spectral lines (but see Section~\ref{sec: soap simulation}). In Section~\ref{sec: inference in lambda}, we use this property to simplify the RV extraction procedure.

\subsection{Is the inference possible?} 
\label{sec: compact support}
The first question at hand is whether a single spectrum provides enough constraints to fit all the parameters of our model.

The spectrum is sampled on a grid with some finite sampling resolution, $\delta \lambda$. A typical sampling resolution of high-resolution spectrographs is  ${\sim}0.1{-}1$ km\,s\textsuperscript{-1}. Due to their compact support, the principal kernels cannot blend values separated by more than $\Delta \lambda_{\max}$, which is on the order of a few tens of km\,s\textsuperscript{-1}. Therefore, to fit a model that consists of about ten principal spectra, the number of parameters we need to infer is
\begin{equation*}
    \frac{\Delta \lambda_{\max}}{\delta \lambda} \times (\text{Number of principal spectra}) \sim 10^3.
\end{equation*}
Since high-resolution spectra typically consist of ${\sim}10^{5}$ sampled wavelengths, a single spectrum should provide enough constraints to make the inference problem feasible, assuming the high-order terms carry the required information content (see discussion in Section~\ref{sec: discussion}).

Without any assumptions regarding the functional shape of the principal kernels,  one can directly fit their values at each sampling point, assuming that
\begin{equation}
    \label{eq: U parameterized}
   {U}^{{\scaleto{(k)}{5pt}}}(\lambda) = 
   \begin{cases}
   \theta^{{\scaleto{(k)}{5pt}}}_\lambda   & |\lambda| \leq \Delta\lambda_{\max}, \\[8pt]
   0                  & \text{else}.
   \end{cases}
\end{equation}
where $\theta^{{\scaleto{(k)}{5pt}}}_\lambda$ is a scalar fitted for every point on the $\lambda$-axis in which the kernel is not zero. Assuming Gaussian white noise, the inference problem can be reduced to least-squares minimisation \citep[e.g.,][]{escude12, zechmeister18, silva22}.

However, while it is established that the inference problem is solvable, the convolution procedure in equation~(\ref{eq: S taylor conv}) makes the solution cumbersome.

\subsection{Simlifying the inference}
\label{sec: inference in lambda}
To simplify the inference procedure, we move to the Fourier domain, where a simple sum of products describes our model (see equation~\ref{eq: S taylor fourier}). In the Fourier domain, the principal kernels are given by
\begin{equation}
    \label{eq: U hat parameterised}
   \tilde{U}^{{\scaleto{(k)}{5pt}}}(\zeta) = {\eta}_\zeta^{{\scaleto{(k)}{5pt}}}.
\end{equation}
However, the number of required parameters and available constraints remained the same as before.

We now utilize the chromatic behaviour of the expansion. We FT a small spectrum segment that starts at some wavelength $\Lambda$,
\begin{equation}
    \label{eq: spec stft}
    \tilde{S}_{\Lambda}(\zeta) = 
    \tilde{U}^{{\scaleto{(0)}{5pt}}}\, \tilde{f}^{{\scaleto{(0)}{5pt}}}_\Lambda(\zeta) + 
    \tilde{U}^{{\scaleto{(1)}{5pt}}}\, \tilde{f}^{{\scaleto{(1)}{5pt}}}_\Lambda(\zeta) + ... \, ,
\end{equation}
where $\tilde{f}^{{\scaleto{(k)}{5pt}}}_\Lambda$ is the FT of the selected segment of the principle spectra. Note that while the principle spectra changed in the transition from equation~(\ref{eq: S taylor fourier}) to~(\ref{eq: spec stft}), their corresponding kernels did not. This is because the principal kernels are insensitive to the wavelength domain in question, by their definition in equation~(\ref{eq: U integral def}). This property simplifies the inference problem: we slice the spectrum into several segments. Since all segments share the same kernels, we have sufficient constraints to fit the Fourier amplitudes of the kernel at each $\zeta$ separately. Once obtained, the derived kernels can be transformed back to the $\lambda$-domain using inverse FT.

This procedure essentially describes a short-time Fourier transform (STFT; e.g., \citealt{grochenig01}, \citealt{muller21}, \citealt{faulhuber22}), which we denote with a tilde sign and $\Lambda$ subscript.
Heuristically, the transform is done as follows:\footnote{
In practice, we derived the STFT using the \texttt{SciPy} 1.10.1 implementation, using a Hann window and 75\% overlap between consecutive segments.} First, we divide each spectrum into $N_\Lambda$ partially overlapping segments of $N_\zeta$ samples each; the segments are then multiplied by some predetermined window function, $w$; finally, the FT of each segment is calculated separately. This procedure turns the one-dimensional spectrum into a two-dimensional matrix. 
Details regarding the STFT procedure, our selected window function, and spectrum segmentation are provided in Appendix~\ref{app: planch}.

\subsection{Modelling the spectrum}
\label{sec: spec modelling}
Using a general prescription of equation~(\ref{eq: U hat parameterised}) for all principal kernels makes interpreting the results harder. Therefore, we use the expansion presented in Section~\ref{sec: Teff expansion} to extract the RV: we explicitly model the zeroth-order term,  $\tilde{S}^{{\scaleto{(0)}{5pt}}}$, and use a general prescription for the high-order ones. The model we fit for the spectrum is given by
\begin{equation} 
\label{eq: rv model stft}
    \tilde{S}_\Lambda(\zeta; \, \Delta v, \pmb{\eta}) = \,\,\tilde{S}_\Lambda^{{\scaleto{(0)}{5pt}}}(\zeta; \Delta v) \, + 
     \sum_{k>0} \eta^{{\scaleto{(k)}{5pt}}}_{\zeta} \tilde{f}^{{\scaleto{(k)}{5pt}}}_\Lambda(\zeta)\, ,
\end{equation}
where $\tilde{S}_\Lambda^{{\scaleto{(0)}{5pt}}}$ is the STFT of the  Doppler-shifted zeroth-order term obtained according to equations~(\ref{eq: S expansion Teff}) and~(\ref{eq: U0}). See Appendix~\ref{app: fit} for a detailed procedure description. 

The number of parameters in this approach can be larger than the minimum required number discussed above. This is because, in this description, we don't only fit the number of points required to characterize the kernels within their compact support --- but rather fit for the entire wavelength segment, which might cover a broader wavelength range compared to the support of the kernels. Other parameterizations may be able to take advantage of the known instrumental line-spread function and further reduce the number of parameters. In this work, we focus on utilizing the representation presented above.

\subsection{The objective function}
So far, we have defined a parameterized model for the spectrum. Now, assume that we were provided with an observed spectrum $D(\lambda)$ from which we wish to infer the model's parameters. 

Assuming the noise is white and Gaussian, the log-likelihood in the $\lambda$-domain is proportional to the negative residual sum of squares,
\begin{equation}
\label{eq: score lambda}
\log\mathcal{L}  + {\rm const}  \propto -\sum_{\lambda}\big|{D}(\lambda)-{S}(\lambda; \, \Delta v, \pmb{\theta}) \big|^2  \, ,
\end{equation}
where $\pmb{\theta}$ is the vector of parameters from equation~(\ref{eq: U parameterized}).
The objective function, which we minimize to derive the set of parameters, is the squared $l_2$-norm of the residual vector.

We now turn to find the equivalent objective function in the STFT domain. To do so, we note that the STFT transform is a tight frame, i.e., the $l_2$ norm of the spectrum in the $\lambda$-domain is proportional to its Frobenius norm in the STFT domain, and the constant of proportion can be derived and accounted for (see Appendix~\ref{app: planch}). Therefore, 
\begin{equation}
\label{eq: score}
\log\mathcal{L}  + {\rm const}  \propto -\sum_{\Lambda \zeta}\big|\tilde{D}_\Lambda(\zeta)-\tilde{S}_\Lambda(\zeta; \, \Delta v,\pmb{\eta}) \big|^2 \equiv - {\rm RSS}\,,
\end{equation}
where $\tilde{D}_\Lambda(\zeta)$ is the STFT of the observed spectrum. Effectively, instead of summing over the squared values of the residuals vector in the $\lambda$ domain, we sum over the squared difference between the STFT matrices of the data and the model, element-wise. 

As equations~(\ref{eq: U hat parameterised}-\ref{eq: rv model stft}) show, the vector of parameters $\pmb{\eta}$ is given per $\zeta$. Therefore, to minimize the objective function, we solve the system of $N_\Lambda$ linear equations for each $\zeta$ separately. The objective function we use is
\begin{equation}
    \label{eq: loss}
    \Delta \log \mathcal{L} \simeq - \frac{ \Delta \,{\rm RSS}}{2 \langle{\sigma}\rangle^{2}}    \, 
\end{equation}
where $\langle{\sigma}\rangle^{2}$ is the variance of the additive Gaussian noise. If the variance is unknown, it could be estimated by $\min\big( { \rm RSS}\big) \big/ N_\lambda$. An outline of the fitting procedure applied in this work is provided in Appendices~\ref{app: fit} and~\ref{app: sufficient stat}.

Equation~(\ref{eq: loss}) implicitly assumes that the error per pixel is constant throughout the spectrum. This is inescapable, as the information regarding the variance of each measurement is not retained in the transition to the STFT domain. However, it is possible to exclude the score of selected sections or weigh their score by the typical inverse variance of the pixel in each segment (see Appendix~\ref{app: fit}).

\section{Numerical experiments}
\label{sec: numerical experiments}
This section demonstrates the application of the STFT-based RV extraction scheme presented above, using simulations based on synthetic spectra. We use the continuum normalized PHOENIX models \citep{husser13}, obtained from the POLLUX\footnote{Publically available via this \hyperlink{http://npollux.lupm.univ-montp2.fr/}{link}.}  database \citep{palacios10}. The models were in the ${\sim}4500{-}7500$ Angstrom wavelength range. 

The first three simulations presented below are simplistic. They are only meant to demonstrate the methodology and illustrate its capabilities in cases where the principal spectra properly account for the physical properties driving the spectral shape variations. In these simulations, we only consider local temperature fluctuation and do not assume any additional physical processes apart from it. We refer to the colder regions on the surface of the simulated star as `spots' for brevity. In Section~\ref{sec: soap simulation}, we apply the method to a realistic simulation.

\begin{figure}
    \centering
        \includegraphics[width=0.975\columnwidth]{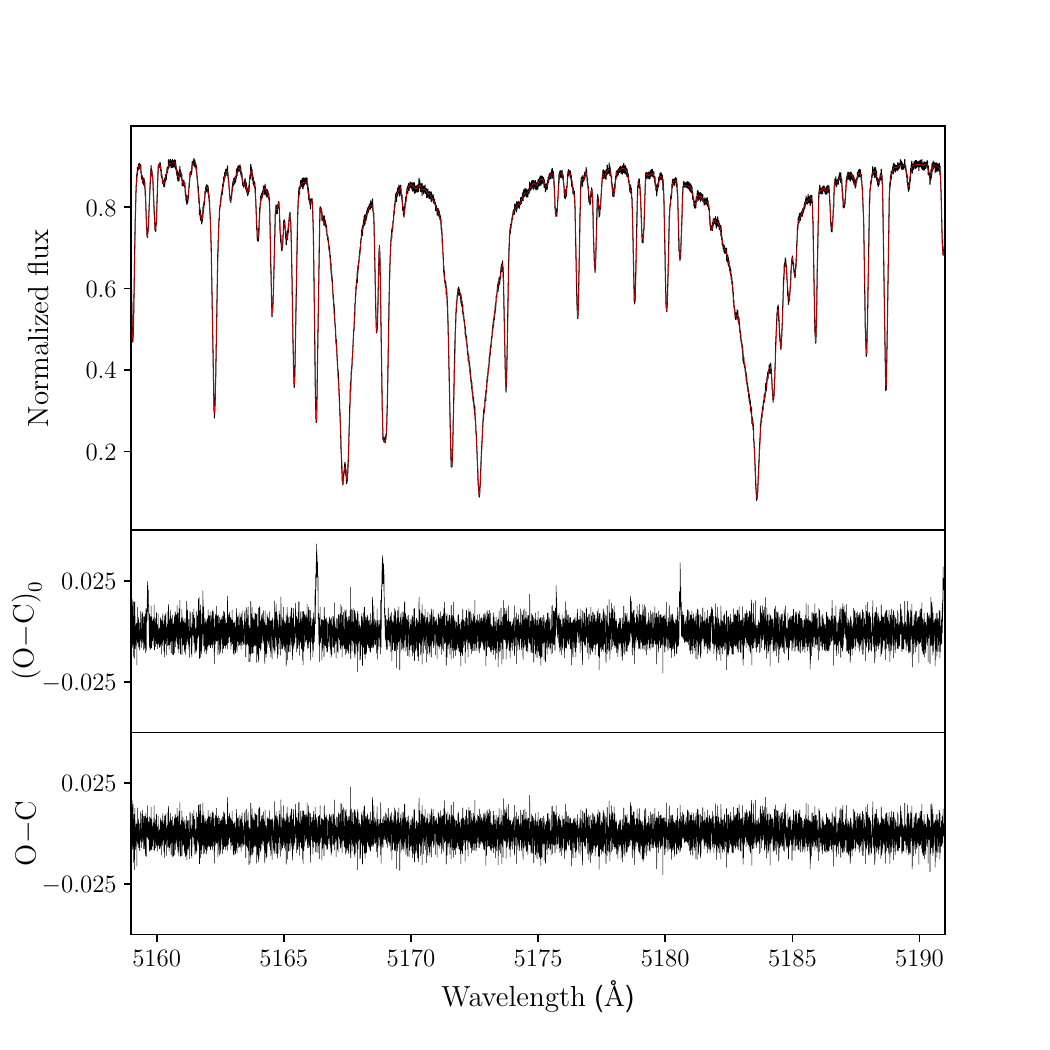} 
        \caption{A fitted spectrum of a Sun-like star with a single dark spot. The spot contribution is taken to be 5\% of the entire flux, its velocity with respect to that of the centre of mass is $400$ m\,s\textsuperscript{-1}, and the temperature difference between the spot and the quiet parts of the star is $200$ K.  Section~\ref{sec: numerical experiments} provides the full simulation description. Top panel --- simulated spectrum (black) and fitted model (red). Middle panel --- the residuals obtained using only the zeroth-order principal spectra. Bottom panel --- residuals from the full model.}
    \label{fig: fit example}
\end{figure}
\begin{figure}
    \centering
        \includegraphics[width=0.975\columnwidth]{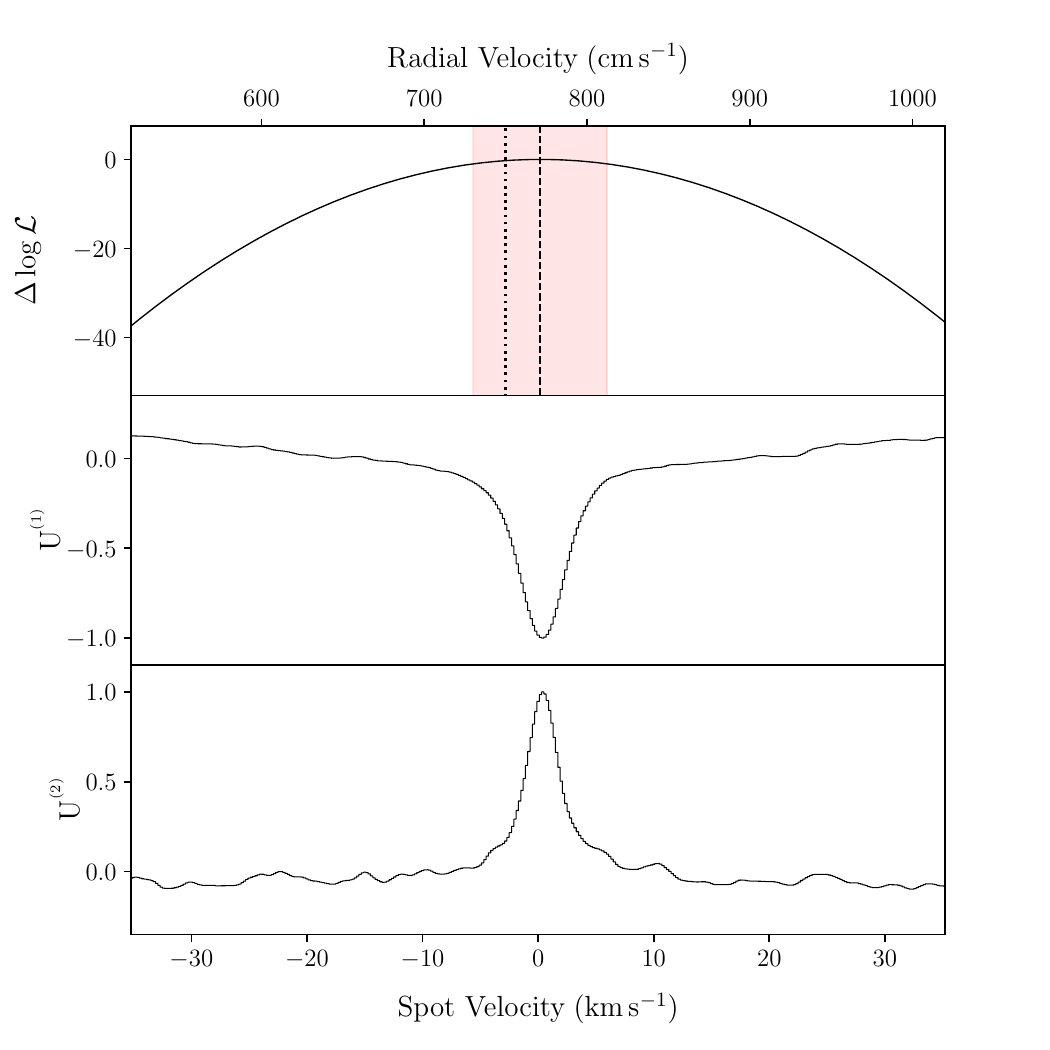} %
        \caption{The log-likelihood and high-order principal kernels for the fit shown in Figure~\ref{fig: fit example}. Top panel --- the log-likelihood versus the RV appears as a solid black line. The vertical dotted and dashed black lines indicate the ground truth and the best-fitting RV values, respectively. The derived one-sigma confidence interval is shown as a red vertical stripe. Middle panel --- the reconstructed first-order principal kernel. Bottom panel --- same as the middle panel, but for the second-order principal kernel. Both kernel plots are normalized, so their maximum absolute value is $1$. }
    \label{fig: fit example 2}
\end{figure} 

\begin{figure}
        \centering        \includegraphics[width=0.975\columnwidth]{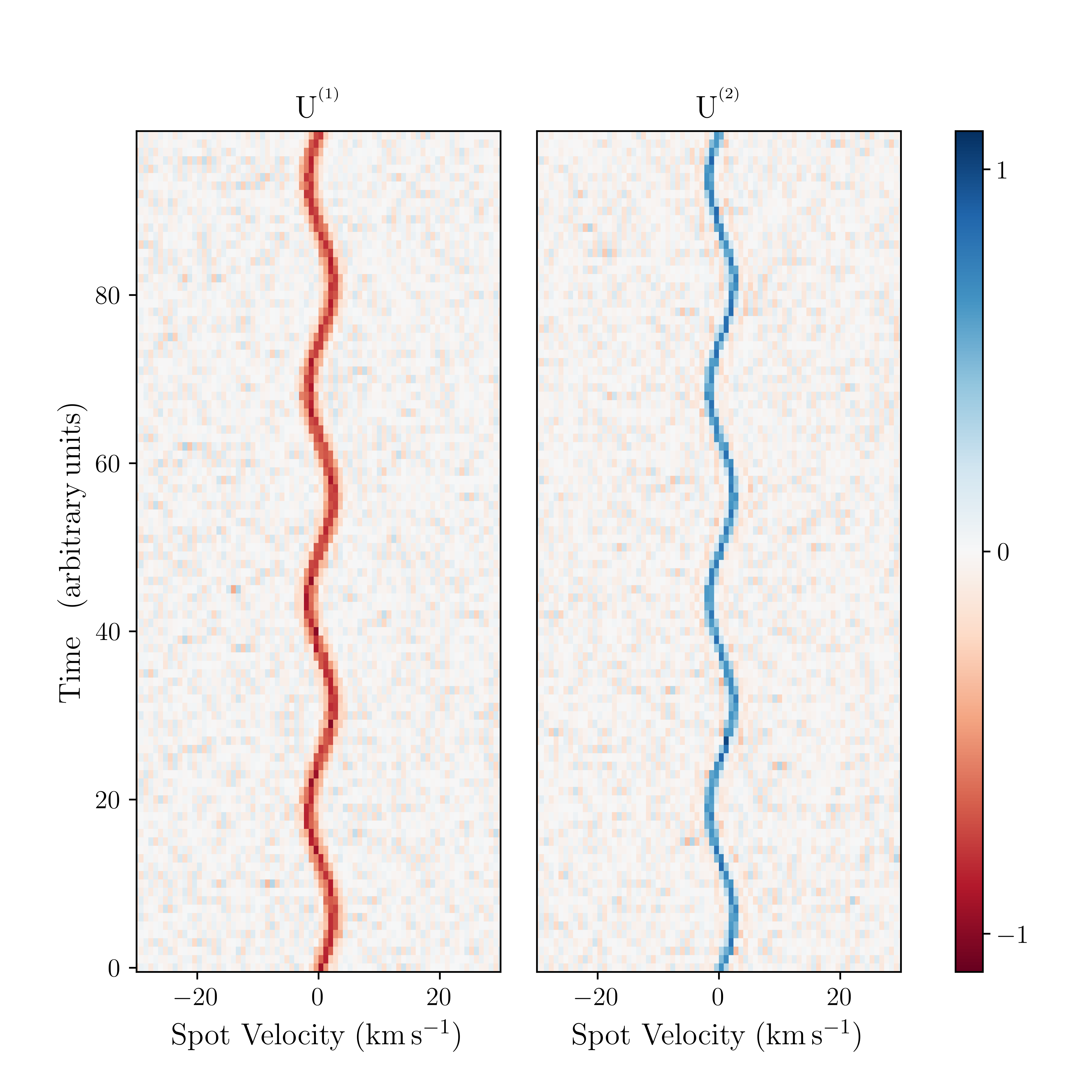} %
        \caption{A river plot of moving spots. The figure illustrates the positions of the high-order principal kernels derived from the simulations described in Section~\ref{sec: moving spot simulation}. Left panel --- the first-order principal kernel corresponding to the photospheric model's first derivative with respect to the effective temperature. Right panel --- the second-order principal kernel corresponding to the model's second effective temperature derivative. The kernels are normalized so that their maximal absolute value is $1$.}
    \label{fig: kernels timeplot}
\end{figure}

\subsection{A single spot}
We start with a simplistic simulation of a star with one spot.  
A continuum-normalized spectrum of a Sun-like star, $S_\star$, is used as the fiducial photospheric model  ($T_{\rm eff}=5{,}800 \, {\rm K}$; $\log g = 4.5$; ${\rm[Fe/H]}=0$; $v \sin i = 5\, {\rm km\, s}^{-1}$). To describe the localized temperature fluctuation, $S_\circ$, we subtract one PHOENIX model from another: a Sun-like photospheric model, with the same stellar properties as the fiducial model but with no rotational broadening, is subtracted from a model with an effective temperature lower by $200$ K.

We Doppler-shifted the stellar model by $7.5$ m\,s\textsuperscript{-1} and the spot spectrum by $400$ m\,s\textsuperscript{-1} with respect to the star's velocity. The two spectra were then co-added, 
\begin{equation}
\label{eq: one spot}
D = (1-\alpha) S_\star + \alpha S_\circ,
\end{equation}
where $\alpha$ represents the fractional contribution of the spot, which we took to be 5\%. The combined spectrum is broadened using a Gaussian profile to simulate an instrument with a spectral resolution, $\mathcal{R} = 200{,}000$ and sampled over a logarithmic grid of $0.2$ km~s\textsuperscript{-1}. Finally, Gaussian white noise was added to the simulated data spectrum, assuming SNR\footnote{The SNR-squared is defined as the ratio between the power in the spectrum and the variance of the additive white noise. We justify this choice in Section~\ref{sec: several spots} and Figure~\ref{fig: snr plot}.} of 20. Figure~\ref{fig: fit example} shows a portion of the resulting spectrum centred around the MgB lines. 

We assume a pixel-dominated noise regime, where the readout, charge transfer and bias contribute most of the noise budget. This simplistic assumption allows a constant-variance additive Gaussian noise model. Depending on the characteristics of the detectors and binning schemes,  we expect it to hold for SNR below ${\sim}{50}$ \citep[e.g.,][]{bouchi09, blackman20, zhao21}. In Section~\ref{section: discussion caveats} and Appendix~\ref{app: fit}, we discuss and show this assumption can be relaxed. 

To derive the RV, we model the spectrum according to equation~(\ref{eq: S expansion Teff}), using the high-order principal spectra according to the $T_{\rm eff}$ expansion shown in Figure~\ref{fig: teff derivatives}. We infer the model's parameters in the STFT domain, as shown in equation~(\ref{eq: rv model stft}), using the loss function given in equation~(\ref{eq: loss}). We scan a range of trial velocities. For each one, we explicitly calculate $\tilde{S}_\Lambda^{{\scaleto{(0)}{5pt}}}$ and analytically derive  the high-order principal kernels, $\pmb{\eta}$, that minimize the loss.  
We used sections of length $2^{10}$ and fractional overlap of 75\% between segments. Since our input is real-valued, this selection yielded $2990$ wavelength segments ($N_\Lambda$), each consisting of $513$ frequencies ($N_\zeta$).

Two `global' parameters, $\Delta v$ and $\alpha$, were fitted to the entire spectrum. Additionally, for each $\zeta$, we fitted two additional parameters -- one per high-order kernel -- using $2990$ constraints. A detailed description of our fitting procedure is provided in Appendix~\ref{app: fit}. The analysis yielded an RV estimate of 
\begin{equation*}
 \hat{v} =  7.71 \pm 0.41 \,\, \text{m\,s\textsuperscript{-1},}
\end{equation*}
which is consistent with the injected velocity of the simulation. The fitted model and its corresponding residuals are shown in Figure~\ref{fig: fit example}. The log-likelihood curve and reconstructed high-order principal kernels are provided in the upper panel of Figure~\ref{fig: fit example 2}. 

We used an explicit inverse discrete FT matrix to reconstruct the shape of the high-order kernels while accounting for the uncertainty in the fitted amplitudes to suppress spurious Fourier components. To do so, we weighted each amplitude by its inverse variance. The recovered kernels, normalised such that their maximum absolute value is unity, appear at their expected position and show an approximately Gaussian profile, as expected. However, we note that the kernels do not drop exactly to zero at high velocities. We ascribe this long-term trend to the fact that the fitted high-order terms are not orthogonal to each other, which might have some implications for the possible application of the method.

\subsection{A moving spot}
\label{sec: moving spot simulation}
To illustrate the role of the high-order kernels, we repeated the single-spot simulation described above. In this exercise, we simulated $100$ spectra; the spectral properties of the star and spot were taken as before but with different velocities. The star was assumed to have zero RV, and the spot of each instance was shifted according to
\begin{equation*}
    v_\circ = 2 \times \sin\big( 2\pi  n/25 \big) \,\,\text{km\,s\textsuperscript{-1}},
\end{equation*}
where $n$ indicates the measurement number (or time, with some arbitrary units). Centre-to-limb variations of the spectrum were neglected in this simulation (but see below).

As before, we model the spectrum according to equation~(\ref{eq: S expansion Teff}), using the high-order principal spectra according to the $T_{\rm eff}$ expansion shown in Figure~\ref{fig: teff derivatives}.  We do not fit the radial velocity of the star, but rather assume it has zero RV and recover the parameters of the high-order principal kernels, $\pmb{\eta}$, by minimising the loss in equation~(\ref{eq: loss}). The recovered kernels are transformed back to the wavelength domain. The time series of recovered kernels reproduce the injected spot velocity, as Figure~\ref{fig: kernels timeplot} demonstrates.

\begin{figure}
        \centering 
        \includegraphics[width=0.975\columnwidth]{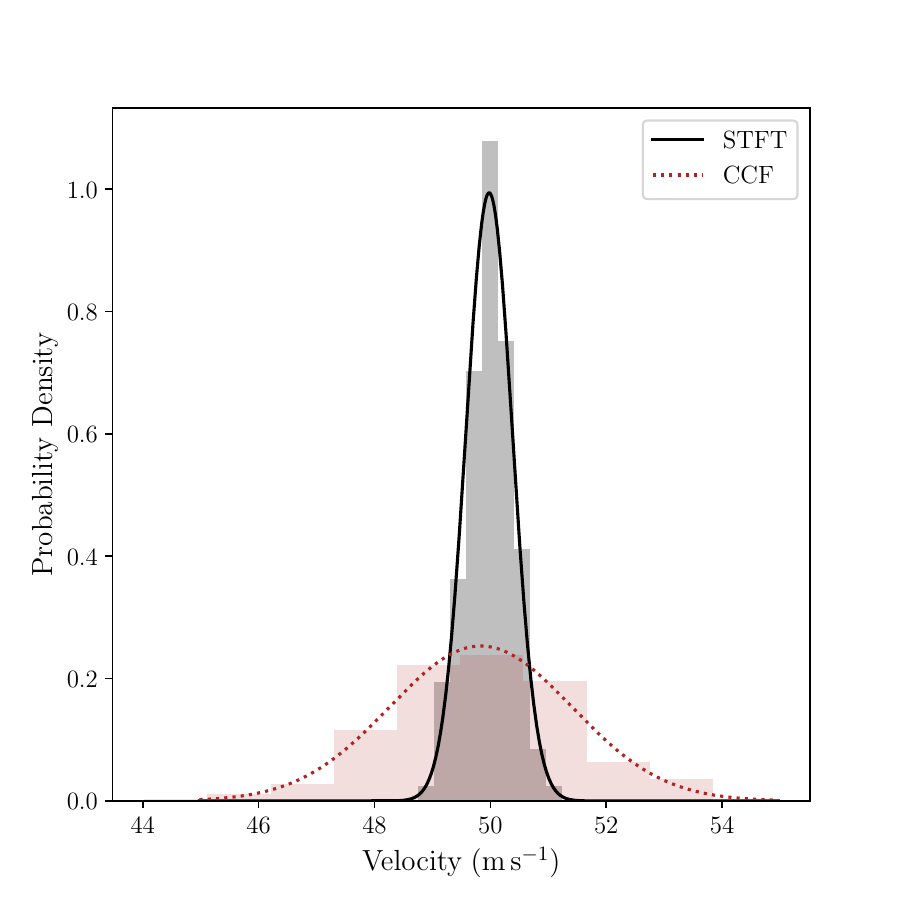} %
        \caption{A Monte-Carlo simulation, extracting the velocities from 100 simulated spectra of a spotted star moving with RV of 50 m\,s\textsuperscript{-1}. See the text for the full details of the simulation. The two histograms show the results of the STFT-based approach to the performance of a standard CCF, showing an improvement by a factor of ${\sim}4$ in the scatter of the derived velocities. Two Gaussian distributions are overlaid, $\mathcal{N}(49.98, 0.40^2)$ and  $\mathcal{N}(49.9, 1.6^2)$, represent the STFT (solid black) and CCF (dotted red) samples, respectively.}
    \label{fig: fit example RVs}
\end{figure}

\begin{figure}
        \centering
        \includegraphics[width=0.975\columnwidth]{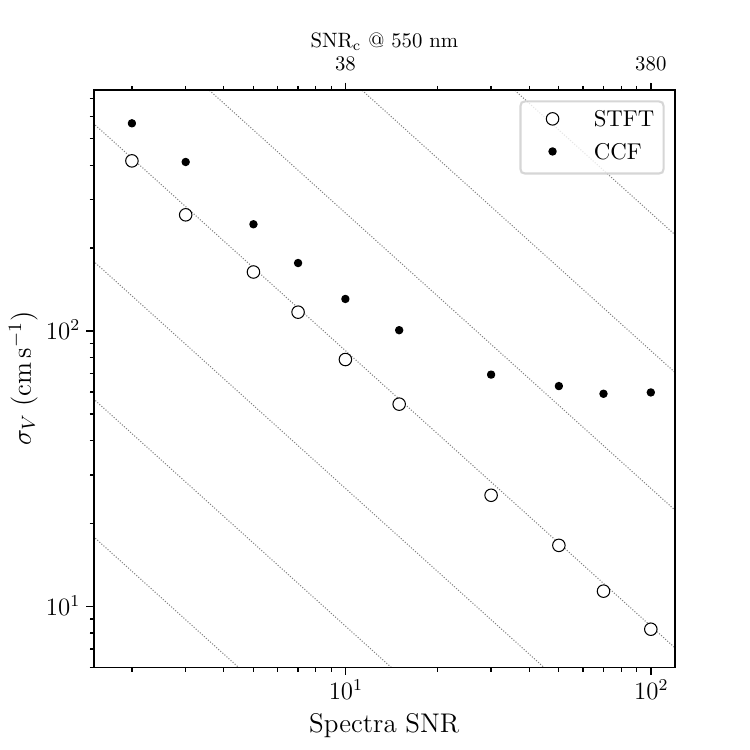} %
        \caption{The scatter in the extracted RV versus the SNR of the simulated spectrum. The scatter of the CCF-derived velocities is shown as black points, and those derived from the STFT method appear as white circles. The grey dotted lines guide the eye to the expected inverse relation. The top x-axis shows the expected continuum-level SNR per pixel at 550 nm (this is a simulation-specific approximated relation; see Appendix E for more details).}
    \label{fig: snr plot}
\end{figure}

\subsection{Several spots}
\label{sec: several spots}
We repeated the experiment above, using similar specifications of the spectral models, but with $5$ spots with random velocities. The spectrum and the spots were added, 
\begin{equation}
D = \big(1-\sum\alpha_i\big) S_\star + \sum\alpha_i S_{\circ, i},
\end{equation}
and broadened by the same assumed instrumental resolution as before. Gaussian noise, equivalent to an SNR of $20$, was added to the data. 
The centre-of-mass velocity was taken to be 50 m\,s\textsuperscript{-1}. The spot velocities were sampled from a Gaussian distribution of zero mean and standard deviations of $200$ m\,s\textsuperscript{-1}, and their fractional contribution, $\alpha_i$, was sampled from a uniform distribution and normalized so that their total contribution, $\alpha = \sum \alpha_i$, was 2.5\%.

We generated $100$ random instances of the configuration above. We derived the RV of each instance twice: First, the RVs were obtained via standard CCF maximisation \citep{tonry79, zucker03}. We cross-correlated\footnote{We calculated the cross-correlation value in the Fourier domain. Optimization was performed using the \texttt{SciPy} optimization and root finding submodule.} the simulated spectra with the quiet star template after applying a Tuckey window to account for edge effects \citep{harris78}. This procedure yielded a sample of RV measurements centred at $49.9$ m\,s\textsuperscript{-1} with a standard deviation of $1.6$ m\,s\textsuperscript{-1},
\begin{equation*}
    \hat{v}_{\textsc{ccf}} \sim \mathcal{N}(49.9, \, 1.6^2)\, .
\end{equation*}
Second, we applied our STFT-based approach, as described above and in Appendix~\ref{app: fit}. This procedure yielded an RV sample centred at $49.98$ m\,s\textsuperscript{-1} with a standard deviation of $0.40$ cm\,s\textsuperscript{-1}, 
\begin{equation*}
    \hat{v}_{\textsc{stft}} \sim \mathcal{N}(49.98, \, 0.40^2)\, .
\end{equation*}
In this numerical experiment, our method improved the precision of the RV estimation by a factor of ${\sim}4$ without impairing the accuracy. Figure~\ref{fig: fit example RVs} shows the resulting RV distributions with the corresponding Gaussian distribution.

Finally, we repeat the Monte Carlo exercise for several different SNR values. If indeed our linearized model can describe the spotted star, we expect that the RV scatter to  be inversely proportional to the SNR of the spectroscopic measurement \citep[e.g.,][]{lovis10}, 
\begin{equation}
\label{eq: snr relation}
    \sigma_V \propto ({\rm SNR})^{-1}.
\end{equation} 
The seed models for the spectra are as before, and so are the size and velocity distributions of the injected spots. We injected $3$ spots with a flux contribution of $0.75$\%. For each SNR value, we repeated the simulation $300$ times. Figure~\ref{fig: snr plot} shows the result of this numerical experiment. The figure shows how the CCF-derived velocities reach a plateau at ${\sim}0.6$ m\,s\textsuperscript{-1} while the STFT-based RV scatter follows the expected relation from equation~(\ref{eq: snr relation}). 

We note that in the simulations we conducted, the ${\rm SNR}$-squared is defined as the proportion of the noise-free model's power, or variance, to that of the additive Gaussian noise. Other alternative definitions might prove more useful or practical in applied stellar spectroscopy. We chose this particular one because it connects the amount of useful information in the data to the amount of noise;\footnote{We point out that our SNR definition should not be confused with the spectroscopic SNR\textsubscript{c}, usually defined in the continuum. For example, SNR of ${\sim}20$ in Figure~\ref{fig: snr plot} is equivalent to  SNR\textsubscript{c} of about ${\sim}80$ (for Sun-like stars). The scaling with RV precision also relies on the instrument's bandpass, sampling, and resolution. Considering realistic specifications of high-resolution instruments, SNR of ${\sim}20$ should yield RV scatter of ${\sim}100$ cm\,s\textsuperscript{-1}. See Appendix~\ref{app: snr} for more details.} this relationship, in turn,  enables the relation in equation~(\ref{eq: snr relation}). The inverse scaling between the RVs and the SNR suggests that our inference procedure performs as expected. 

The value of the RV scatter in Figure~\ref{fig: snr plot} depends on specific simulation details, such as the instrumental resolution, sampling rate and wavelength range. Of course, if the noise were induced by some different source, for which the high-order terms cannot account, the STFT-based method would have encountered a noise floor of its own. This is demonstrated in the following subsection.

\begin{figure}
        \centering 
        \includegraphics[width=0.975\columnwidth]{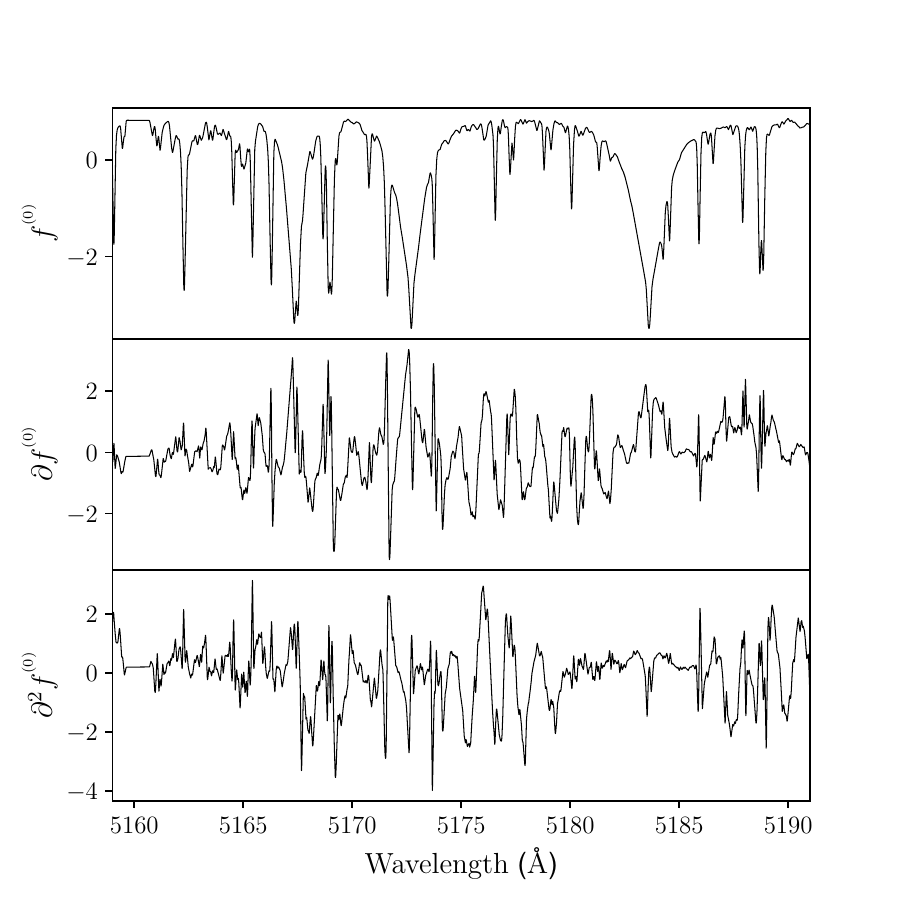} %
        \caption{Mean-subtracted and normalized sections of principal spectra used for the linearized photospheric models in Section~\ref{sec: soap simulation}. Top panel ---  the zeroth-order principal spectra, $f^{{\scaleto{(0)}{5pt}}}$, a Solar spectrum, taken where the cosine of the heliocentric angle, $\cos \mu$, equals $0.3$.  Middle panel --- the first-order principal spectra, $f^{{\scaleto{(1)}{5pt}}}$, taken as the derivative of the model with respect to $\cos \mu$. Bottom panel --- the second-order principal spectra, $f^{{\scaleto{(2)}{5pt}}}$, taken as the second derivative of the model with respect to $\cos \mu$.}
    \label{fig: clv derivatives}
\end{figure}

\begin{figure}
        \centering 
        \includegraphics[width=0.975\columnwidth]{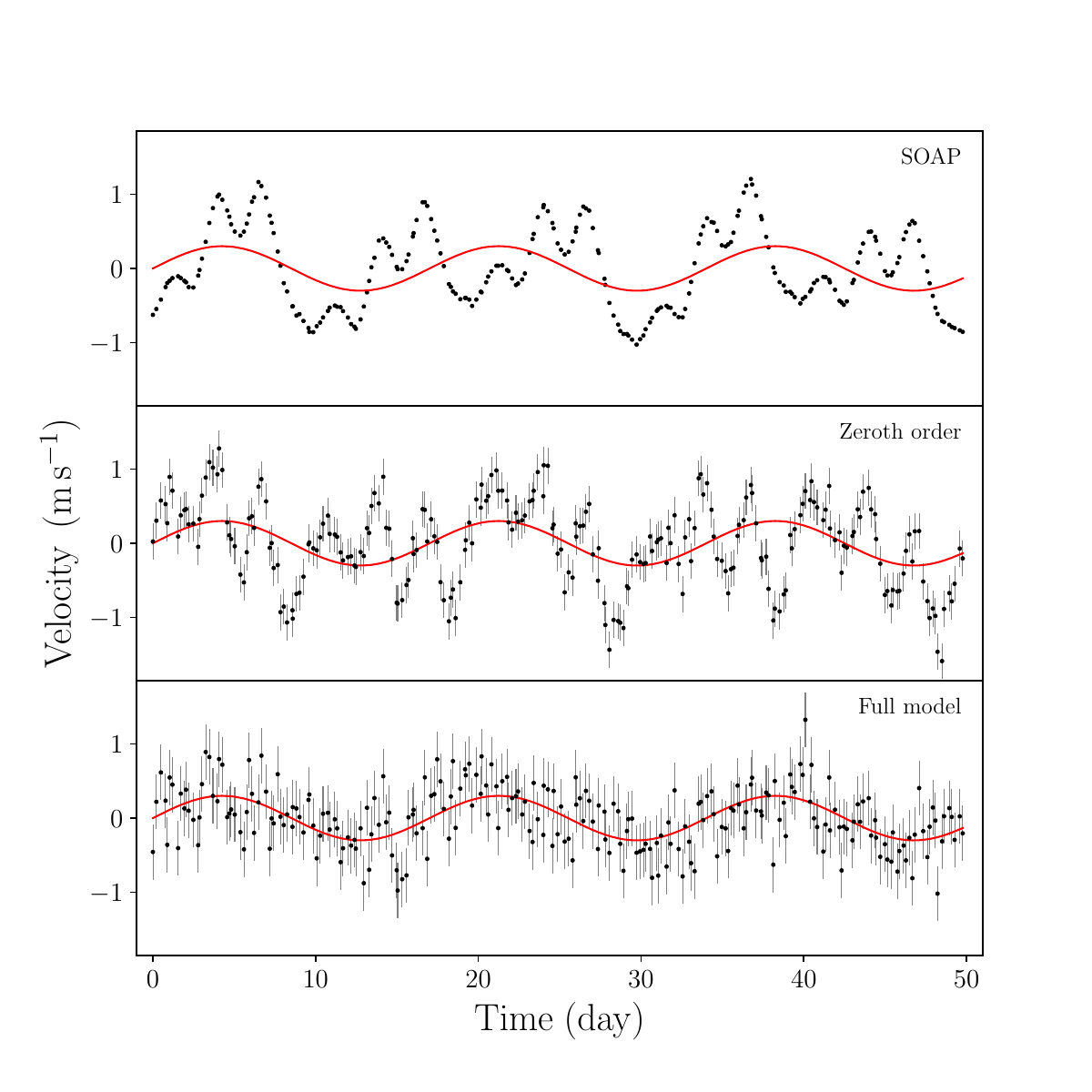} %
        \caption{Top panel --- the mean-subtracted default RVs provided with the SOAP simulation, added with the simulated planetary-like periodic Doppler shift (solid red line). Middle panel --- a time series of the velocities derived using only the zeroth-order term, without the centre-to-limb principal spectra. 
        Bottom panel --- a time series of the velocities derived using the full model, using five centre-to-limb principal spectra.}
    \label{fig: SOAP RV}
\end{figure}

\subsection{SOAP simulations}
\label{sec: soap simulation}
To further demonstrate the method, we test it against a realistic simulated data set produced using the GPU-accelerated version of Spot Oscillation And Planet software (SOAP-GPU; \citealt{zhao23}). The simulations were produced by Zhao \& Dumusque (private communication).

\subsubsection{Properties of the simulated dataset}
A spectrum of the quiet Sun served as the seed model of the simulation. Spots and faculae were generated using PHOENIX models with effective temperatures of $5015$ and  $6028$ $^\circ {\rm K}$, respectively. { The spot sizes and spot-to-faculae size ratios were randomly drawn according to \citet[][table 1 therein]{borgniet15}. The latitude distribution of the injected active regions follows the butterfly diagram, and their longitudes are separated by 180 degrees, reproducing the active longitudes (\citealt{borgniet15}). Active regions were injected into the simulated spectra while accounting for convective blueshift inhibition and the dependence of the line bisector on the centre-to-limb angle, $\mu$.  There is no size evolution of the simulated active regions.}

There are 249 simulated spectra in the synthetic dataset, taken continuously over 50 days with an average time interval of 5 hours between each observation. { The simulated star has a rotation period of 10 days; on average, it has 10 active regions on its surface, with a median spot size of ${\sim} 50$ and an interquartile range of ${\sim}15{-}250$ millionths of a Solar hemisphere. The expected value for the active region filling factor is ${\sim}1$ per cent, considering the spot and faculae-to-spot size distributions.} We periodically shifted the spectrum to simulate the reflex motion due to a planet, 
\begin{equation*}
    v = 30 \times \sin\bigg[ \frac{2\pi\, (t-t_0)}{17~{\rm day}} \bigg] \,\,\text{cm\,s\textsuperscript{-1}},
\end{equation*}
where $t_0$ is the epoch of the first simulated spectrum. The spectrum was sampled with a logarithmic grid of 0.8 km~s\textsuperscript{-1}, similar to what is provided by many high-resolution spectrographs. Gaussian white noise was added to the simulations, assuming an SNR of 100.

\subsubsection{Centre-to-limb principal spectra}
Studies and observations suggest that the suppression of granular blueshift in magnetised regions of the Sun's photosphere is responsible for most of the parasitic activity-induced RV signal \citep[e.g.,][]{darvins81, haywood16}. This effect has been found to impact the detectability of exoplanets \citep[e.g.,][]{reiners16, meunier17, celga19}, and its spectral signature has been associated with $\mu$ angle at which the active region is located \citep[e.g.][]{bauer18}. Even if one assumes that convective blueshift inhibition is a perfectly `grey' effect, namely, it only affects the local velocity field without changing the spectral shape of the active region, it will still have a spectral signature in the disk-integrated spectrum of the star. This is due to the variation in the stellar spectrum from the star's centre to its limb.

We can account for these variations as small deviations from our analysis's standard rotational broadening kernel. To produce the required centre-to-limb principal spectra, we use the IRSOL Solar atlas,\footnote{Available via this \href{https://www.irsol.usi.ch/data-archive/clv-ss3/}{link}.} compiled by \citet{ramelly17}. 
The catalogue provides a high-resolution spectrum of the Sun, taken at different heliocentric angles, $\mu$. Figure~\ref{fig: clv derivatives} presents the Solar spectrum taken at $\cos \mu = 0.3$ along with the first two derivatives. The calculation and role of the derivatives are identical to the effective temperature derivatives used in our naive temperature fluctuation example in Section~\ref{sec: Teff expansion}.

\subsubsection{Analysis and results}
To derive the RV, we model the spectrum similarly to what is described in equation~(\ref{eq: S expansion Teff}), but using the first five derivatives with respect to $\cos \mu$ as the high-order principle spectra, the first two of which are presented in the middle and bottom panels of Figure~\ref{fig: clv derivatives}. For the zeroth-order term, we used the same Sun-like PHOENIX model used in our previous simulations, shown in the top panel of Figure~\ref{fig: teff derivatives}. We used shorter sections of length $2^{6}$ and the same fractional overlap of 75\%  between segments. Similar results were obtained with section lengths of up to $2^8$. The data and code underlying this numerical experiment are available online.\footnote{The data underlying Figure~\ref{fig: SOAP RV} is available via a \href{https://github.com/saharsh1/STFT-RV-example}{GitHub repository}.}

The results of our analysis are presented in Figure~\ref{fig: SOAP RV}. The bottom panel shows our full-fitted model's extracted velocity and estimated uncertainty, with a solid red line depicting the expected RV modulation due to the simulation reflex motion of the star. The middle panel shows the extracted RVs using only the zeroth-order Sun-like PHOENIX model. The top panel of Figure~\ref{fig: SOAP RV} shows, for reference, the RV estimates obtained by cross-correlating the simulated spectrum with a predetermined mask (\citealt{zhao23}). These velocities, provided as part of the simulation data products, were added with the assumed value of the RV with which we shifted the data.  The figure illustrates that our method substantially reduced the effect of stellar activity. The resulting scatter around the injected Doppler shift is ${\sim}34$ cm\,s\textsuperscript{-1}. Considering the coarser sampling and shorter wavelength coverage of the SOAP simulations, the expected shot-noise limit RV scatter is ${\sim}35$ cm\,s\textsuperscript{-1} (see Appendix~\ref{app: snr}), consistent with the results of the simulation. Nevertheless, some small residual correlations can be seen in the derived velocities. This is probably because the high-order principal spectra used in our calculation do not perfectly describe the activity-induced variations in the spectrum.

\section{Discussion}
\label{sec: discussion}
In this paper, we assumed that the emitted spectrum per unit area of the star is describable by a linear combination of {\textit known} principal spectra. Building on this assumption, we showed that the integrated spectrum attains a simple form: a sum of principal convolution kernels that operate on the principal spectra.

The information content of the spectrum can be efficiently encapsulated in the kernels. The principal kernels are a \textit{sufficient statistic}, and, under our assumptions, no other statistic provides additional information on the RV value or the deformations of the line profile. We showed how to recover the principal kernels using least squares fitting in the STFT domain and showed the mathematical equivalence of our procedure to least-squares fitting in the wavelength domain.

We first demonstrated the method's performance on some simplistic simulations of a star with temperature fluctuations on its surface. This simulation serves as a proxy for a star-spotted spectrum while maintaining the fidelity of the principal kernels we currently have at hand. Our naive simulations are simplistic but conceptually similar to modern codes that produce realistic spectra (e.g., \citealt{dumusque14, zhao23, palumbo22}). This is because they represent, at least approximately, the integrated spectrum of many parcels on the star's surface.

As a more realistic demonstration, we analysed a set of SOAP simulations. To do so, we derived the principal spectra associated with the Sun's centre-to-limb variation. As Figure~\ref{fig: SOAP RV} shows, our analysis can mitigate the effects of stellar activity. However, as discussed above, it is likely that the high-order terms used in the process only do not fully account for the spectral variability. Therefore, we expect that the performance of this particular setup will change with spot coverage and temperature contrast variations of the simulated spectrum (see Section~\ref{section: discussion caveats}).

\subsection{Towards extreme precision RV}
The scope of this work is limited to developing and presenting the formalism underlying the RV extraction method. Some of the examples provided above are simplistic, as we aim to illustrate the role of the different components in the factorisation and the method's potential.

Moving forward, the most crucial step would be confronting real solar or stellar spectra. To do so, we must obtain realistic estimates, either theoretical or empirical, of the principal spectra. Various aspects of stellar spectra can be assumed to vary smoothly with a small number of parameters. The effective temperature was discussed as a fiducial example in this work, but other effects are likely to be described by a similar prescription. For example, \citet{pietrow22} studied the centre-to-limb variation in the line profile; \citet{al-moulla22} linked the local magnetic fields, driving convective blue shift inhibition, with the formation temperature of each part of the spectral lines;  and, \citet{palumbo22} developed an efficient empirical model to describe the effect of granulation. These studies and others can serve as a starting point in obtaining linearized prescriptions of stellar emission patterns. 

Not all effects that are imprinted in the measured spectrum are associated with the astrophysical object we observe. For example, atmospheric transmission, line-spread-function variations (e.g., \citealt{gilbertson22, allart22}), and calibration drift also produce chromatic effects on the stellar spectrum.  At first glance, some of these effects might be hard to approach using a linearized approach. However, recent studies have shown significant success in mitigating the effect of telluric lines using linear methods such as principal component analysis (e.g.,  \citealt{bedell19}, \citealt{cretignier21}, and references therein) or as a smooth function of a few parameters \citep{ivanova23}. Our factorisation can therefore help account for telluric contamination. The expansion approach shown in Section~\ref{sec: Teff expansion} is particularly suited to mitigate the impact of micro-telluric lines on the RV estimate \citep{cunha14, wang22, latouf22}. 

Provided with linearised models, our method could serve as a step towards achieving the community's RV precision goals in the coming years. Building on the results from various recent studies, principal spectra can be derived, enabling our formalism to account for various aspects of stellar- or instrumental-induced chromatic variations.  Recent studies of the Sun as a star (e.g., \citealt{al-moulla22b}; \citealt{lin22}; \citealt{ellwarth23}) provide an opportunity to empirically obtain the shape of the principal spectra and apply this knowledge to the analysis of other stars. However, deriving the corresponding principal spectra requires substantial study and analysis and therefore deferred to follow-up work.

\subsection{Increased search volume}
An additional consequence of this work is that it demonstrates how precision RV measurements can be achieved with moderate SNR compared to what is currently required. 

The primary goal of the Terra Hunting Experiment\footnote{\href{https://www.terrahunting.org/}{www.terrahunting.org}} \citep{hall18} is to monitor a sample of at least $20$ G or K dwarf stars, with accuracy sufficient to detect Earth-analogues. However, as recently stated by \citet{gupta23}, the yield of this survey relies on the mitigation of known noise sources. For most methodologies, mitigating these noise sources requires a high SNR per frequency element, raising the cost in terms of the required observing time. The possible causes for this are diverse. For example, activity indicators that are derived from a narrow portion of the spectrum constrain the SNR requirements per pixel. Indicators extracted from the CCF also require a high SNR because the spectral information is diluted in this process.

Our methods can potentially mitigate the contribution of parasitic RV signals induced by various stellar or instrumental effects. As Figure \ref{fig: snr plot} shows, this will allow RV scatter to approach the shot-noise limit. The method presented in this work does not rely on a specific set of lines or a small spectral region, but rather analyses all spectral lines and their differential reactions to changes in the physical conditions on the stellar surface. Our approach boils down the information content to a relatively small number of parameters while retaining a sufficient statistic (see Appendix~\ref{app: sufficient stat}). As a result, information is not lost in the process, as opposed to cross-correlation-based techniques.

Therefore, the method presented in this work provides an opportunity to increase the viable sample size of spectroscopic campaigns. This could be achieved either by observing fainter targets, reaching more targets per observing run, or by constraining the orbital solutions to sufficient accuracy with fewer observations.

\subsection{Caveats}
\label{section: discussion caveats}
The methodology and examples in this work are based on certain assumptions about spectrum modeling and measurement noise properties. In this section, we highlight a few related key points and challenges.

\subsubsection{Approximated photospheric models}
Figure~\ref{fig: teff derivatives} indicates that, within the parameter space considered in our simulations, temperature fluctuations primarily impact only a few specific spectral lines, leaving most of the spectrum largely unaffected. Conversely, the centre-to-limb terms presented in Figure~\ref{fig: clv derivatives} appear to have a broader impact. This property hints that the parasitic RV signal induced by stellar activity is caused predominantly by convective blueshift inhibition, even though temperature differences between active regions can exceed a thousand Kelvin. This is consistent with the literature \citep{haywood16} and with our analysis of SOAP simulations.

Another factor that can affect the effectiveness of the procedure is the approximation efficiency. The test cases shown in Section~\ref{sec: several spots} suggest that two temperature derivatives can account for local temperature fluctuations (or inaccuracy of the temperature estimate) of up to ${\sim}200$ Kelvin. However, as illustrated in Figure~\ref{fig: teff map}, certain spectral lines might change drastically with the effective temperatures, potentially appearing or vanishing entirely. Hence, for large temperature changes, the procedure might necessitate additional high-order temperature derivatives.

Lastly, we note that the dominant expansion parameters could vary based on spectral type, stellar activity, or observed wavelength range. Optimal leading terms for a Sun-like star might differ from those required to analyse spectra of M-dwarf or fast-rotating massive stars. Furthermore, in some cases, the limitations of the modelling could affect efforts to obtain reliable spectra and derivatives, as noted in \citet{olander21}.

\subsubsection{Noise model}
In this work, we assumed that all pixels are subjected to the same noise distribution. Yet, in practice, both the intensity of stars and the instrumental sensitivity are a function of the wavelength. 

This limitation comes naturally from Fourier domain analysis. When transitioning from equation~(\ref{eq: score lambda}) to~(\ref{eq: score}), our independent variable changes from pixel to frequency. Although this limitation is a natural consequence of our inference method, it can be mitigated. Our inference technique employs the STFT, where individual spectrum segments undergo their own Fourier transforms. In our examples, our analysis was performed based on nm or sub-nm segments. Therefore, one simple way to adjust our procedure to account for non-uniform variance is by assigning inverse-variance weights to the segments, as detailed in Appendix~\ref{app: fit}.

\subsection{Prospects for future work}

As stated above, this paper is limited in its scope to developing and introducing our formalism. The applicability of our method will depend on our ability to provide realistic linearised models for the observed spectrum. The expertise required to utilise this technique is distributed in the community, spanning from theoretical modelling to instrumental design. While we fully intend to apply our method to real data to Solar spectroscopic measurements and test our results on RV standards (e.g., \citealt{zhao22}), it is also evident that knowledge transfer is much needed to achieve optimal results. In conclusion, we briefly outline the main avenues for development.

The analysis of real spectroscopic data sets presents additional inherent challenges. The main points to which we believe our method might be sensitive include the consistency of the blaze and continuum-level rectification; line spread function variation, either in wavelength or time; and telluric or Solar contamination. Similar challenges were recently addressed by other studies, for example, by \citet{hirano20, lienhard22} with promising results. The immediate course of action towards the application of our method on stellar spectra is by studying and analysing Solar data \citep[e.g.,][]{cameron19, dumusque21,lin22}. The results from the SOAP simulation analysis shown in Section~\ref{sec: soap simulation} suggest that our method carries the potential to mitigate Solar variability. Therefore, Solar spectra can serve as a benchmark, isolating the impact of calibration and environmental effects on the inference method.

From a signal-processing point of view, our method could be extended in several directions. First, in this work, we only consider the analysis of a single spectrum. However, the principle kernels describe the star's physical state at a specific epoch. Measuring the temporal evolution of the high-order kernels could serve as an activity indicator of sorts, probing the temporal evolution of the stellar surface (see Figure~\ref{fig: kernels timeplot}, for example). Second, we explore the utilisation of bilinear optimisation algorithms, like generalisations of singular value decomposition (e.g., \citealt{golub96, edelman20}) to simultaneously constrain the principal spectra and kernels for a set of many (${\sim}100$) spectra of a single target. This would relax, to some extent, the need to have a detailed description of the processes that govern the spectral shape.

Finally, while this work primarily addressed the goal of achieving extreme RV precision, we believe that other applications are feasible. For example, spectral disentanglement of spectroscopic binaries, tomographic studies of stellar surfaces, and studies of late-type stars with poorly constrained models.

\section*{Acknowledgements}
We are indebted to the two anonymous referees for their rigorous review and constructive recommendations. Their suggestions significantly improved this work.
This research benefited from the SOAP-GPU simulations Yinan Zhao and Xavier Dumusque generated. We are grateful for their contribution, which significantly improved this work. This study greatly benefited from the lively discussions and interactions the fifth Extreme Precision Radial Velocity (EPRV) Conference enabled. We particularly thank Khaled Al-Moulla, Michael Cretignier,  Xavier Dumusque, Eric B. Ford, Nathan Hara and Suvrath Mahadevan for the illuminating discussions. We also thank Avraham Binnenfeld, Sagi Ben-Ami, Dotan Gazith, Oryna Ivashtenko, Eran Ofek and Shay Zucker for their helpful comments and advice.

This study used the photospheric models obtained from the \hyperlink{http://pollux.oreme.org}{POLLUX database} operated at LUPM with the support of the PNPS and INSU. The solar atlas used in this work has been provided in electronic form by IRSOL as a compilation by \citet{ramelly17}. We \texttt{matplotlib} \citep{Hunter_2007}; \texttt{NumPy} \citep{Numpy_2006, Numpy_2011}; and \texttt{SciPy} \citep{2020SciPy-NMeth}. 

The research of SS is supported by the Benoziyo Prize Postdoctoral Fellowship. BZ is supported by a research grant from the Willner Family Leadership Institute for the Weizmann Institute of Science, a research grant from the Center for New Scientists at the Weizmann Institute of Science and a research grant from the Ruth and Herman Albert Scholarship Program for New Scientists.

\section*{Data Availability}
The photospheric models used in the simulations are available via the \hyperlink{http://pollux.oreme.org}{POLLUX database}. The Solar centre-to-limb spectra are available via the \href{https://www.irsol.usi.ch/data-archive/clv-ss3/}{IRSOL Solar atlas} webpage. The data underlying Figure~\ref{fig: SOAP RV} are available via a \href{https://github.com/saharsh1/STFT-RV-example}{GitHub repository}.



\bibliographystyle{mnras}
\bibliography{main}



\appendix

\section{Broadening kernels}
\label{app: broadening kernels}
In Section~\ref{sec: Teff expansion}, we use the zeroth order spectrum to derive the RV of the star and treat the high-order terms as a perturbation to it. The general form of the zeroth-order principal kernel is provided in equation~(\ref{eq: U0}), in terms of  the star's RV, its rotational velocity projection on our line of sight, its linear limb-darkening coefficient and the instrumental resolution ($\Delta v$, $v_{\rm rot}$, $\epsilon$ and $\mathcal{R}$, respectively). 

The explicit analytical form of the kernels used in equation~(\ref{eq: U0}) is 
\begin{equation}
\label{eq: app kernels}
\begin{aligned}
\tilde{K}_{\rm inst}(\zeta) & \propto \exp{\bigg(- \frac{\pi^2 }{4 \ln{2}}} \frac{c^2}{\mathcal{R}^2} \,  \zeta^2 \bigg)
\text{\,\,\, and}\\[8pt]
\tilde{K}_{\rm rot}(\zeta) & \propto \frac{6}{x}\frac{1-\epsilon}{3-\epsilon}  J_1 \big(x\big) + \frac{6}{x^3}\frac{\epsilon}{3-\epsilon}\big( \sin x - x \cos x\big), \\[8pt]
\end{aligned}
\end{equation}
where $J_1$ represents the first order Bessel function of the first kind and, for convenience, we denoted $x\equiv2\pi  \zeta v_{\rm rot}$. We also note that as $\tilde{K}_{\rm rot}$ approaches unity as $x$ goes to $0$. For more details, see \citet{diaz07} and references therein.

\section{Short time Fourier transform}
\label{app: planch}
STFT is a procedure in which one transforms a one-dimensional signal into a two-dimensional function, describing the signal's frequency content at different segments. In our case, the one-dimensional input signal is the spectrum $S(\lambda)$. 

The procedure is as follows: The spectrum is divided into segments of equal length. We then multiply each segment by some predetermined window function, $w$. Finally, we calculate the Fourier transform of each segment separately. The transform yields a function of two independent variables: the FT variable, $\zeta$, and the characteristic wavelength of each segment, $\Lambda$. STFT therefore maps the spectrum --- a vector in $\mathbb{R}^{\textsc{n}_\lambda}$ --- into a two-dimensional diagram --- a matrix in $\mathbb{C}^{{\textsc{n}_\zeta\times \textsc{n}_\Lambda}}$.

Formally, for some window function $w$, the STFT of the $\Lambda$\textsuperscript{th} segment of the spectrum is given by
\begin{equation}
    \tilde{S}_\Lambda(\zeta) = \sum_\lambda{ S(\lambda) w(\lambda - \Lambda) e^{-2
    \pi i\zeta \lambda / N_\zeta }}\, ,
\end{equation}
where, as the equation shows, $\Lambda$ is the reference zero-point of the segment.
The shape of the window function and the overlap between consecutive segments determine the properties of the resulting transform. Several valid choices for the window function exist, and we refer the readers to the literature for information (for example, \citealt{SASPWEB2011}). In this work, however, we consider only the Hann window, 
\begin{equation}
    w(\lambda) = 
    \begin{cases}
    \frac{1}{2} -\frac{1}{2} \, \cos \bigg( \frac{2\pi \lambda}{\Delta \Lambda}  \bigg) \quad & \text{if} \, \, 0 < \lambda < \Delta \Lambda , \\[8pt]
    0 & \text{else}.
    \end{cases}
\end{equation}
where $\Delta \Lambda$ is the segment's length.

If the overlap fraction between consecutive segments in the decomposition is exactly 50\%, the Hann window fulfils the constant overlap-add criterion,
\begin{equation}
    \sum_{n\in\mathbb{Z}} w\bigg(\lambda -  n \frac{\Delta\Lambda}{2}\bigg) = 1.
\end{equation}
This is usually a desirable property. If fulfilled, it follows that the sum of all STFT sections coincides with the FT of the signal, 
\begin{equation}
     \tilde{S}(\zeta) = \sum_\Lambda \tilde{S}_\Lambda(\zeta)   \,,
\end{equation}
where $\Lambda = n \cdot \Delta \Lambda /2$, for some integer $n$. 

However, for our needs, this property is insufficient. For equation~(\ref{eq: loss}) to hold, the transform must conserve the energy in the signal, at least up to some known constant factor. However, for a Hann window with 50\% overlap 
\begin{equation}
    \label{eq: app ineq 1}
    \sum_{n\in\mathbb{Z}} w^2\bigg(\lambda -  n \frac{\Delta\Lambda}{2}\bigg) \neq \text{const},
\end{equation}
therefore, the energy is not conserved. Explicitly, this means that 
\begin{equation}
     \sum_\zeta|\tilde{S}(\zeta)|^2 \neq \sum_{\zeta, \Lambda} |\tilde{S}_\Lambda(\zeta)|^2   \,,
\end{equation}
and the ratio between the terms on the two sides of the inequality is not a constant number.
The inequality above, unless remedied, makes the transition from equation~(\ref{eq: score lambda}) to equation~(\ref{eq: score}) not valid.

One possible solution to this situation is normalising the window function by the factor in equation~(\ref{eq: app ineq 1}). The \texttt{SciPy} STFT implementation supports this normalization. A more direct approach would be selecting the overlap fraction differently. By using a Hann window with an overlap fraction of 75\%, we get that 
\begin{equation}
    \sum_{n\in\mathbb{Z}} w^2\bigg(\lambda -  n \frac{\Delta\Lambda}{4}\bigg) = \frac{3}{2}\,.
\end{equation}
where now we take $\Lambda = n \cdot \Delta \Lambda /4$.
In this case, energy is conserved up to a known factor of three halves. Hence
\begin{equation}
     \sum_\zeta|\tilde{S}(\zeta)|^2 = \frac{2}{3} \sum_{\zeta, \Lambda} |\tilde{S}_\Lambda(\zeta)|^2   \,.
\end{equation}
We note that, in this setup, the constant overlap-add requirement is also met (up to a known factor of two). In summary, using a Hann window and a fractional overlap of 75\%, we ensured that the STFT is a \textit{tight frame}: the transformed signal conserves the energy up to a known constant, and a perfect reconstruction is possible.

Elaborated discussions regarding this Plancherel-like for the continuous STFT can be found in the recent pedagogical summary by \citet[][Corollary 2.8, therein]{faulhuber22}. For the case of discrete sampling, see \citet{griffin84} and \citet{muller21}.

\section{fit}
\label{app: fit}

In this appendix, we describe in more detail the fitting procedure used for the numerical simulations in Section~\ref{sec: numerical experiments}. We aim to describe the likelihood calculation of some trial velocity, $\Delta v$. 

First, we note that shifting the spectrum or the model in the Fourier domain is simple multiplication by a complex phase. However, shifting in the STFT domain should be done cautiously because we must account for the shift's effect on the transformed model's window function. Unaccounted for, this will induce a modelling error proportional to  ${\sim} \Delta v N_\Lambda   (\Delta \Lambda)^{-1}$. To avoid this issue, we first shift by the required velocity and only then apply the STFT. Since we have several principle spectra and only one observed spectrum, we shift the observed data by $-\Delta v$, which is more efficient than shifting all the principle spectra by $\Delta v$. The likelihood function in equation~(\ref{eq: score lambda}) is unaffected by this choice.

The procedure we take is as follows:  First, we shift the observed spectrum by $-\Delta v$. We then estimate the flux contribution of the spot, $\hat{\alpha}$, by a simple linear fit to $\tilde{S}^{{\scaleto{(0)}{5pt}}}$. We then apply STFT to the residuals, forming an STFT matrix ${\bf y}$ of dimensions $N_\zeta \times N_\Lambda$ (see Appendix~\ref{app: planch}). For each $\zeta$, this matrix is equivalent to a vector of size $N_\Lambda$, 
\begin{equation}
    \mathbfit{y}_\zeta \equiv
\begin{pmatrix}
    \vert \\
     \tilde{D}_\Lambda(\zeta) \\
    \vert 
\end{pmatrix}
- \hat{\alpha}
\begin{pmatrix}
    \vert \\
   \tilde{S}^{{\scaleto{(0)}{5pt}}}_\Lambda  \\
    \vert 
\end{pmatrix} \, ,
\end{equation}
where $\hat{D}_\Lambda$ represents the STFT decomposition of the observed spectrum shifted by $-\Delta v$. We assumed in the process that a reliable estimate of the instrumental resolution and rotational broadening profile is available. Extending this procedure to account for additional parameters is straightforward.

The second step is fitting the high-order terms to the matrix ${\bf y}$ in the STFT domain.
The design matrix ${\bf X}$ contains the STFT of the decomposed high-order kernels. Therefore, it is a complex three-dimensional matrix in $\mathbb{C}^{{\textsc{n}_\zeta\times \textsc{n}_\Lambda\times \textsc{n}_k}}$, where the size along each dimension are determined by the STFT decomposition ($N_\zeta$ and $N_\Lambda$) and the number of high order terms, $N_k$. Because the fit is done per $\zeta$, we can consider separately the values taken along the first dimension,
 \begin{equation}
{\bf X}_\zeta \equiv
\begin{pmatrix}
    \vert & & \vert \\
     \tilde{f}^{{\scaleto{(1)}{5pt}}}_\Lambda(\zeta)  & \cdots  & \tilde{f}^{{\scaleto{(N_k)}{5pt}}}_\Lambda(\zeta)  \\
    \vert &   &\vert 
\end{pmatrix} \,.
\end{equation}

The standard solution to the normal equations gives the estimated coefficients for each $\zeta$,
\begin{equation}
    \label{eq: eta hat app}
    {\hat{\pmb{\eta}}}_\zeta = \big( {\bf X}^{H}_\zeta {\bf X}_\zeta\big)^{-1} \, {\bf  X}^{H}_\zeta \,\mathbfit{y}_\zeta \,.
\end{equation}
Finally, we calculate the residual sum of squares from equation~(\ref{eq: score})
\begin{equation} 
\label{eq: RSS app}
 {\rm RSS} = \sum_{ \zeta}\big|\big| {\bf y}_\zeta - {\bf X}_\zeta  \cdot \hat{\pmb \eta}_\zeta \big|\big|^2_2 \, .
\end{equation}
The RSS value above is calculated for a grid of trial velocities. This procedure yields a log-likelihood curve via equation~(\ref{eq: loss}) from which we estimate the RV and its confidence interval, as demonstrated in Figure~\ref{fig: fit example 2}.

The equations above assume that the uncertainty over the entire spectrum is approximately constant. As discussed in Section~\ref{section: discussion caveats}, this is probably not the case in reality. One simple way to address it is by weighing each section $\Lambda$ by its inverse variance. To do so, we define the weights to be inversely proportional to the typical variance of each segment, namely, $w_\Lambda \propto \langle \sigma \rangle^{-2}_\Lambda$. Now, we can rewrite equation~(\ref{eq: eta hat app}) 
\begin{equation}
    \label{eq: eta hat app w}
    {\hat{\pmb{\eta}}}_\zeta = \big( {\bf X}^{H}_\zeta {\bf W} \,  {\bf X}_\zeta\big)^{-1} \, {\bf  X}^{H}_\zeta {\bf W} \, \,\mathbfit{y}_\zeta \,.
\end{equation}
where {\bf W} is a diagonal matrix of size $N_\Lambda\times N_\Lambda$, in which the $i$\textsuperscript{th} diagonal term contains the weight of the corresponding wavelength segment. To calculate the segment-weighted score, we rewrite equation~(\ref{eq: score}),
\begin{equation} 
\label{eq: RSS app}
\Delta \log \mathcal{L} \simeq -\frac{1}{2}\sum_{ \zeta} \big({\bf y}_\zeta - {\bf X}_\zeta  \cdot \hat{\pmb \eta}_\zeta \big)^{H} \,{\bf W}  \,\big({\bf y}_\zeta - {\bf X}_\zeta  \cdot \hat{\pmb \eta}_\zeta \big).
\end{equation}

\section{Sufficient statistic}
\label{app: sufficient stat}
Finally, we show that the maximum-likelihood estimator from equation~(\ref{eq: eta hat app}), $\pmb{\eta}_{\zeta}$, together with the  ${\bf X}_\zeta^{H} {\bf X}_\zeta$ matrix,  form a \textit{sufficient statistic} \citep{degroot05}. To do so, we need to show that we can calculate the likelihood of any parameter vector, ${\pmb v}_\zeta$, up to a multiplicative constant using $\pmb{\eta}_{\zeta}$ and ${\bf X}_\zeta^{H} {\bf X}_\zeta$  alone, without explicitly using  $\mathbfit{y}_\zeta$. Equivalently, we can require that RSS of each $\zeta$ term in the sum of equation~(\ref{eq: RSS app}) can be calculated up to an additive constant using only the suggested statistic. 

To show that this requirement holds, we note that each term in the RSS sum can be written as
\begin{equation}
{\rm RSS}_{\zeta} = \big|\big| {\bf X}_\zeta  \cdot {\pmb v}_\zeta \big|\big|^2_2  - 2 \Re \big( {\pmb v}_\zeta^{H} {\bf X}^{H}_\zeta  {\bf y}_\zeta \big) + \big|\big| {\bf y}_\zeta \big|\big|^2_2\,.
\end{equation}
The first term depends only on the ${\bf X}_\zeta^{H} {\bf X}_\zeta$ and the trial parameter vector but not on the data; therefore meets our requirement. The last term is an additive constant that depends only on the data and, therefore, meets our requirements. Still, we need to show that the middle term can be rewritten without explicit dependence on ${\bf y}_\zeta$. To do so, we note that by using equation~(\ref{eq: eta hat app}), we can rewrite the mixed term as  
\begin{equation}
  {\pmb v}_\zeta^{H} {\bf X}^{H}_\zeta  {\bf y}_\zeta =    {\pmb v}_\zeta^{H} \big({\bf X}^{H}_\zeta {\bf X}_\zeta) \, {\hat{\pmb{\eta}}}_\zeta \, .
\end{equation}
  
Using the two equations above, we can obtain, up to a constant, the likelihood of every trial parameter vector ${\pmb v}_\zeta$ while using only $\pmb{\eta}_{\zeta}$ and ${\bf X}_\zeta$, as required. The assumptions underlying our derivation are that the star's spectrum can be represented as a linear combination of the principal spectra and that the noise is approximately Gaussian.

\section{Practical aspects of SNR estimstes}
\label{app: snr}
The SNR definition used in the simulations is applicable since the spectrum is known. However, this definition might be inconvenient when organizing an observing campaign. A more practical approach is to use the continuum-level signal-to-noise ratio, ${\rm SNR}_{\rm c}$, estimated at some fixed wavelength. 
This appendix provides an approximated relation between the two definitions. 

To do so, we first note that the RV variance, $\sigma_{\rm V}^2$, is proportional to the typical width of the spectral lines \citep{lovis10} and inversely proportional to the effective number of pixels \citep{bouchi09}. Therefore,  using the scaling shown in Figure~\ref{fig: snr plot}, the inverse-proportion relation in equation~(\ref{eq: snr relation}) becomes
\begin{equation}
    \label{eq: SNR scaling}
    \sigma_{\rm V} \approx 0.3 \,\bigg( \frac{30}{\rm SNR} \bigg) \sqrt{\bigg( \frac{2\cdot10^5}{\mathcal{R}} \bigg) \bigg( \frac{\delta \lambda}{200 \, {\rm m\,s}^{-1}} \bigg) \bigg( \frac{300 \, {\rm nm}}{\rm BP}\bigg)} \,\,{\rm m\,s}^{-1}\,,
\end{equation}
where $\mathcal{R}$, $\delta \lambda$ and ${\rm BP}$ and the instrumental resolution, average sampling interval, and the width of the spectral bandpass, respectively, and we have assumed that the instrumental resolution dominates the line broadening. 
For example, the SOAP simulations presented in Section~\ref{sec: soap simulation} were produced assuming SNR of 100, resolution of $115000$, sampling rate of $800$ m\,s\textsuperscript{-1} and bandpass of $140$ nm. The predicted scatter in this case is ${\sim}0.35$ m\,s\textsuperscript{-1}, consistent with the simulation results. 

We now consider the scaling relation derived by \citet[][]{petersburg20} for EXPRES,
\begin{equation}
\sigma_{\rm V} \approx 0.3 \, \bigg(\frac{250}{\rm SNR}_{\rm c}\bigg)  \,\,{\rm m\,s}^{-1}.
\end{equation}
The observed target was 51 Peg; its properties are similar to the synthetic spectra used in our simulations, and the continuum-level SNR was estimated at $550$ nm.
The velocities were obtained using a $250$ nm spectral range, sampling rate of ${\sim}550$ m\,s\textsuperscript{-1}, and spectral resolution of $137500$, yielding
\begin{equation}
{\rm SNR}_{\rm c} \approx 3.8 \times {\rm SNR}.    
\end{equation}
This approximated scaling relation is not general and applies to slow-rotating Sun-like stars observed by instruments similar to EXPRES. Careful considerations of the target stars, instruments, and observing conditions are warranted to obtain more accurate estimates.


\bsp	
\label{lastpage}
\end{document}